\documentclass[12pt]{article}
\usepackage[T1]{fontenc}
\usepackage{a4wide}
\usepackage{psfrag}
\usepackage[dvips]{graphicx}
\usepackage{amssymb}
\usepackage{geometry}
\usepackage{subfigure}
\makeatletter

\long\def\@makecaption#1#2{%
  \vskip\abovecaptionskip
  \sbox\@tempboxa{#1#2}%
  \ifdim \wd\@tempboxa >\hsize
    #1: #2\par
  \else
    \global \@minipagefalse
    \hb@xt@\hsize{\hfil\box\@tempboxa\hfil}%
  \fi
  \vskip\belowcaptionskip}

\renewcommand\theequation{\hbox{\normalsize\arabic{section}.\arabic{equation}}}
\@addtoreset{equation}{section}
\renewcommand\thefigure{\hbox{\normalsize\arabic{section}.\arabic{figure}}}
\@addtoreset{figure}{section}

\@addtoreset{table}{section}                                                    

\newcommand{\LyX}{L\kern-.1667em\lower.25em\hbox{Y}\kern-.125emX\spacefactor1000}

\newcommand{\eqalign}[1]{
\null \,\vcenter {\openup \jot \ialign {\strut \hfil $\displaystyle {
##}$&$\displaystyle {{}##}$\hfil \crcr #1\crcr }}\,}
\newcommand{\be}{\begin{equation}}
\newcommand{\ee}{\end{equation}}
\newcommand{\ba}{\begin{array}}
\newcommand{\ea}{\end{array}}
\newcommand{\bea}{\begin{eqnarray}}
\newcommand{\eea}{\end{eqnarray}}
\newcommand{\pa}{\partial}

\newcommand{\AmS}{{\protect\the\textfont2
  A\kern-.1667em\lower.5
ex\hbox{M}\kern-.125emS}}
\newcommand{\lyxaddress}[1]{
  \par {\raggedright #1 
  \vspace{1.4em}
  \noindent\par}}

\makeatother

\begin{document}

\title{\begin{flushright}
{\small ITP -- Budapest Report 570}
\end{flushright}\vspace{1cm}
\textbf{\large Boundary states and finite size effects in
sine-Gordon model\\ with Neumann boundary condition}\large }

\author{Z.~Bajnok\protect\( \protect \), L.~Palla\protect\( \protect \)\thanks{
Corresponding author's e-mail: palla@ludens.elte.hu
}~~ and G.~Tak\'{a}cs\protect\( \protect \)}

\maketitle

\lyxaddress{\centering \protect\( \protect \)\emph{Institute for Theoretical Physics
}\\
\emph{E\"{o}tv\"{o}s University} \emph{}\\
\emph{H-1117 Budapest, P\'{a}zm\'{a}ny P. s\'{e}t\'{a}ny 1 A-\'{e}p, Hungary}}

\begin{abstract}
The sine-Gordon model with Neumann boundary condition is
investigated. Using the bootstrap principle the spectrum of boundary
bound states is established. Somewhat surprisingly it is found that 
 Coleman-Thun diagrams and bound state creation may
coexist. A framework to describe finite size effects in boundary
integrable theories is developed and used together with the truncated
conformal space approach to confirm the bound states and reflection
factors derived by bootstrap.
\end{abstract}

{\par\centering PACS codes: 64.60.Fr, 11.10.Kk   
\par}

\clearpage

\section{Introduction}

In 2 dimensions certain integrable quantum field theories can be
restricted to the $x\le 0$ half line by imposing suitable boundary
conditions without destroying their integrability \cite{GZ}. In
addition to their theoretical  interest, these models have
important physical applications in various  impurity problems 
(for a review see \cite{rev}). 

An example is provided by the sine-Gordon model: as was argued in
\cite{GZ}, its boundary version:
 \be
S=\int\limits_{-\infty}^\infty dt\int\limits_{-\infty}^0 dx{\cal
L}_{SG}-\int\limits_{-\infty}^{\infty}dt V_B(\Phi_B),\qquad\quad 
 {\cal L}_{SG}=\frac{1}{2}(\pa_\mu\Phi )^2-\frac{m^2}{\beta^2}(1-\cos
(\beta\Phi )),
\ee
(where $\Phi (x,t)$ is a scalar field, $\beta $ is a real dimensionless
coupling and $\Phi_B (t)=\Phi (x,t)\vert_{x=0}$) preserves the
integrability of the bulk if the boundary potential is chosen as
\[
V_B(\Phi_B)=-M_0\cos\left(\frac{\beta}{2}(\Phi_B-\phi_0)\right),
\]
where $M_0$ and $\phi_0$ are free parameters. A novel feature of the
boundary sine-Gordon model (BSG) is the appearance of a complicated
spectrum of boundary bound states (BBS) in addition to the well known
bulk ones \cite{GZ}-\cite{patr}, \cite{skod}. The complete spectrum 
of these bound states and
a full explanation of all the poles are known only for BSG with Dirichlet
boundary conditions (which corresponds to taking
$M_0\rightarrow\infty$, $\Phi_B(t)\equiv\phi_0$)
 \cite{patr}.

In this paper we investigate the bound state spectrum in sine-Gordon
model with Neumann boundary condition (SGN). This boundary condition
is in a sense the opposite limit to Dirichlet ($M_0=0$, thus $\phi_0$
becoming irrelevant) and has interesting properties as the
non conservation of topological charge. Furthermore any BSG model,
which is not Dirichlet, behaves in the ultraviolet limit as if it had
Neumann boundary condition. 

We determine the spectrum of boundary bound states by using the
bootstrap principle and give in fact an inductive self consistent
proof of their existence. We also give the explanation of all the
poles in the various reflection factors in terms of these bound states
and Coleman-Thun diagrams \cite{CT}. The importance of Coleman-Thun
diagrams in the context of boundary bootstrap was first emphasized in
\cite{bCT}. An interesting feature of SGN is that we find instances
when Coleman-Thun diagrams and bound state creation coexist (a similar
phenomenon was previously observed in the case of boundary Yang-Lee
theory \cite{bCT}). To discover this one has to compute and compare
the residues of Coleman-Thun diagrams and the reflection factors, thus
one has to go beyond the usual argument that checks only the existence
of a diagram with the required pole.

Once we know the spectrum of SGN on the infinite half line the next
step is to inquire how it is changing when the model is restricted to
a finite line segment $0\le x\le L$ with suitable boundary
conditions. We develop a transfer matrix formalism and use it to
discuss the description of boundary bound states in this setting in some detail. Since
the resulting finite volume spectra depend on the reflection factors,
they provide an ideal laboratory to check these reflection factors, if
we can measure the spectra independently. 

To this end we develop the truncated conformal space approach
(TCSA) \cite{YZ} to SGN. Determining the finite volume spectra
numerically by using TCSA we are able to compare them to the
theoretical predictions, and the complete agreement we find gives a
strong evidence for the correctness of the various bound states and
reflection factors.

The paper is organized as follows: in the next section we describe
some classical solutions of SGN and show that the semi-classical
quantization of the boundary breathers explains a part of the poles
in the breathers reflection factors on non excited boundary. In
section 3, using the bootstrap principle, we derive the spectrum of
boundary bound states and the associated reflection factors. Section 4
is devoted to the discussion of finite size effects in a large volume
$L$, with some non trivial boundary conditions at the ends of $L$. In
section 5, as the main step of TCSA, the Hamiltonian of BSG on $0\le
x\le L$ is described as that
of a bulk and boundary perturbed free boson with appropriate boundary conditions. 
We compare in detail the TCSA data and the theoretical predictions
 in section 6. We make our conclusions in section 7. The paper is
closed by three appendices: in Appendix A we give in details  the proof of the
existence of the boundary bound states. In Appendix B we explain the
poles of the soliton and breather reflection factors. Finally, in
Appendix C we review those aspects of boundary $c=1$ theories that
are necessary to set up our TCSA.

\section{A few classical solutions and the semiclassical spectrum of boundary breathers in SGN}

In boundary sine-Gordon theory with Neumann boundary condition 
(SGN) the boundary potential is absent: $M_0=0$.
As a result, $\Phi (x,t)$ satisfies free boundary condition at
$x=0$: $\partial_x\Phi (x,t)\vert_{x=0}=0$. This boundary condition
preserves the $\Phi\leftrightarrow -\Phi$ charge conjugation symmetry
(${\bf C}$ symmetry) of the bulk theory, but violates the conservation
of topological charge. This can be seen already on the simplest
classical solution of SGN, describing the scattering of a classical
soliton on the boundary. This solution, $\tilde{\Phi}_s(x,t)$, is
obtained by restricting to the $x\le 0$ half line the $\Phi_{SA}(x,t)$
soliton anti soliton solution of the bulk theory:
\be
\tilde{\Phi}_s(x,t)\equiv \Phi_{SA}(x,t)=
\frac{4}{\beta}{\rm arctan}\Bigl[ 
\frac{\sinh (umt/\sqrt{1-u^2})}
{u\cosh (mx/\sqrt{1-u^2})}\Bigr],\quad 
{\rm for}
\quad -\infty <x\leq 0.
\ee
(Since $\partial_x\Phi_{SA} (x,t)\vert_{x=0}=0$ it indeed satisfies
the Neumann boundary condition). This solution exhibits that the
incident soliton reflects as an anti soliton from the Neumann boundary,
i.e. topological charge changes by two units in this scattering.  

Another observation that plays an
important role in the sequel is that 
the \lq standing' breather solution of 
the bulk SG,  
oscillating with period $\tau$ around $x=0$: 
\be
\label{lelegzo}
\Phi_\tau (x,t)=\frac{4}{\beta}{\rm arctan}\Bigl[ \sqrt{(\tilde{\tau})^2-1}
\frac{\sin (mt/\tilde{\tau})}
{\cosh (mx\sqrt{1-(\tilde{\tau})^{-2}})}\Bigr],\qquad 
\tilde{\tau}=\frac{m\tau}{2\pi},
\ee
also satisfies 
$\pa_x\Phi_\tau (x,t)\vert_{x=0}=0$. Therefore the classical breather 
solution, bound to the wall at $x=0$  is given by 
\be
\label{blelegzo}
\tilde{\Phi}_\tau (x,t)=\Phi_\tau (x,t),\qquad {\rm for}
\qquad -\infty <x\leq 0.
\ee

In SGN 
\lq boundary dependent' simple poles are found 
in the reflection factors of the various breathers \cite{gosh} that may 
describe boundary bound states (BBS). In the reflection factor of the $n$-th 
breather, $B^n$, this pole is located at the 
$\theta =inp\pi /2$ \footnote{The parameter $p$ is determined by the sine-Gordon coupling constant
as $p=\beta^2/(8\pi -\beta^2)$.}   
value of the
purely imaginary rapidity, so that if it really corresponds to a BBS, then
the energy of this state above the ground state is 
\be\label{eegye}
e_n-e_0=M\sin
(np\pi ).
\ee 
In the form of two lemmas sufficient conditions were given in \cite{patr} 
that guarantee that a simple pole in a reflection factor 
cannot be explained by the Coleman-Thun 
mechanism \cite{CT}, i.e. that it describes a bound state. However, these conditions are 
not satisfied for any of the poles above, thus the question whether they correspond
 to BBS remains open. Furthermore, a Coleman-Thun explanation of 
a subset of the poles
is given by the solitonic version of diagram (c) on
Fig.(\ref{fig:bound}), 
when the  index $n$ of the breather satisfies 
$n>1/(2p)$. (This is discussed in detail in the next section).
We clarify the status of the poles for $n<1/(2p)$ below by showing that 
the energies associated to them as hypothetical BBS match exactly with the 
 spectrum of bound states obtained by semi-classical quantization 
from the classical boundary breather solutions of SGN
(\ref{lelegzo}-\ref{blelegzo}). 
We achieve this by 
adapting to this problem the semi-classical derivation of the breather 
spectrum in the bulk sine-Gordon theory as given in the classic 
paper \cite{dhn}.   
  
The semi-classical (WKB) quantization of any periodic classical solution,
$\phi_{cl}$ in a field theory   
can be summarized by the equations:
\be
\label{WKB1}
-\frac{\pa}{\pa\tau}\Bigl[ S_{cl}(\phi_{cl})+S_{ct}(\phi_{cl})
-\sum\limits_{i=0}^{\infty}(n_i+\frac{1}{2})h\nu_i(\phi_{cl})\Bigr] =E,
\ee
\be
\label{WKB2}
W_{\{ n_i\}}(E)=2K\pi\hbar ,\qquad K\quad {\rm integer},
\ee
where
\be
\label{WKB3}
W_{\{ n_i\}}(E)=S_{cl}(\phi_{cl})+S_{ct}(\phi_{cl})+E\tau (\phi_{cl})
-\sum\limits_{i=0}^{\infty}(n_i+\frac{1}{2})h\nu_i(\phi_{cl}).
\ee
Here $\tau(\phi_{cl})$ denotes the period of the classical solution, 
$\nu_i(\phi_{cl})$ stand for its stability frequencies (that characterize 
the quasi periodicity $\xi_i(x,t+\tau )={\rm e}^{i\nu_i}\xi_i(x,t)$ of the
solutions of
$
\Bigl[ -\frac{\pa^2}{\pa t^2}+\frac{\pa^2}{\pa x^2}
-\bigl(\frac{\pa^2 U}{\pa \phi^2}\bigr)_{\phi_{cl}}\Bigr]
\xi_i(x,t)=0$), 
and $n_i$ are a set of non negative integers. Furthermore the subscript 
\lq ct' stands for counter term contributions, whose effect is to cancel 
divergences in the infinite sums. Eq.(\ref{WKB1}) picks out the appropriate
classical periodic solution for a given $E$ and a given set of integers 
$n_i$. Then those values of $E$ which satisfy Eq.(\ref{WKB2}-\ref{WKB3}), 
give the renormalized bound state energies. 
 
With the Neumann boundary condition the boundary potential is absent, and 
as ${\cal L}_{SG}(\Phi_\tau)$ is symmetric for $x\mapsto -x$, we 
easily obtain
\be
S_{cl}(\tilde{\Phi}_\tau )=\frac{16\pi}{\beta^2}
\left({\rm arccos}\left(\frac{1}{\tilde{\tau}}\right)-\sqrt{(\tilde{\tau})^2-1}\right),
\ee
which is just {\it half} of the corresponding expression in the bulk theory. 
The stability frequencies for $\tilde{\Phi}_\tau$ are the {\it same} as for
$\Phi_\tau$, but the $\tilde{\xi}(x,t)$ fluctuations must also satisfy
$\pa_x \tilde{\xi}(x,t)\vert_{x=0}=0$. Therefore only  half of 
the bulk stability fluctuations appear in the Neumann problem, namely those that are 
even under $x\mapsto -x$, (and the same applies to the fluctuations 
contributing to the vacuum energy $E_{vac}$ appearing in the counter term). 
Thus effectively we must take $\sum\tilde{\nu}_i=\frac{1}{2}\sum\nu_i$. 
To obtain the basic boundary bound states we set all $n_i=0$, and using 
the explicit form of the counter-terms and the sums over $\nu_i$ as
given in \cite{dhn} we find finally
\be
S_{ct}(\tilde{\Phi}_\tau ) -\sum\frac{1}{2}\tilde{\nu}_i
=-\frac{\beta^2}{8\pi}S_{cl}(\tilde{\Phi}_\tau ).
\ee  
Using these results in Eq.(\ref{WKB1}, \ref{WKB3}) leads to
\be
\tilde{E}=\frac{m}{\pi p}\frac{\sqrt{(\tilde{\tau})^2-1}}{\tilde{\tau}},
\qquad W(\tilde{E})=\frac{2}{p}{\rm arcsin}
\left(\frac{\tilde{E}p\pi}{m}\right).
\ee
 As a result the quantization
condition, $W(\tilde{E}_K)=2\pi K$, gives
\be 
\tilde{E}_K=M\sin (Kp\pi ),
\ee
where $M=\frac{m}{p\pi}$ is the semi-classical soliton mass. Clearly
this reproduces (\ref{eegye}). 
Since 
$\tau (\tilde{E}_K)\sim\frac{1}{\cos (Kp\pi)}$ the range of $K$-s where 
bound states exist (i.e where $\tau (\tilde{E}_K)<\infty$)  is $K<1/(2p)$.

\section{Boundary bound state spectrum from bootstrap principle}

In this Section we determine the spectrum of boundary excitations,
the related soliton and breather reflection factors and show how their
poles can be explained in terms of on-shell diagrams. We start with
a summary of the bulk scattering properties and then review the result
of Ghoshal and Zamolodchikov concerning the reflection factors of
the soliton and breathers on the non-excited boundary. From their
pole structure and the bootstrap equation we conjecture the minimal
spectrum of the excited boundary states (or with other words boundary
bound states), and prove in Appendix A that they can be created
by successive soliton absorptions on the wall at purely imaginary
rapidity. Having determined the reflection factors we explain all
their poles in terms of on-shell diagrams which correspond either
to some Coleman-Thun mechanism or creation of boundary bound states.
In this study we rely heavily on the machinery worked out in
\cite{patr} for the Dirichlet case.
However, in contrast to the Dirichlet case we find instances
when a Coleman-Thun type diagram happens to give a pole of the same order 
as the reflection amplitude has, but with wrong residue, thus leaving
the possibility of excited boundary state creation at the same time.

\subsection{Bulk scattering properties}

\begin{figure}~~~~~\subfigure[Breather fusion]{\resizebox*{!}{5cm}{\includegraphics{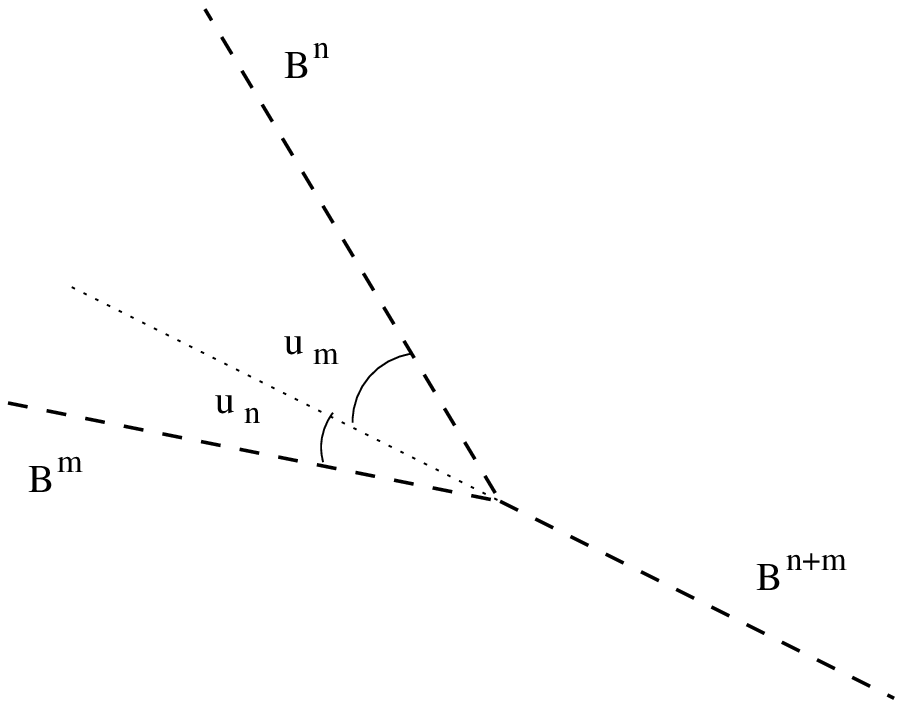}}} 
~~~~~~~~~~~~~~~~~~~~~~~~~~\subfigure[Soliton and anti soliton fuse to a
breather]{\resizebox*{!}{5cm}{\includegraphics{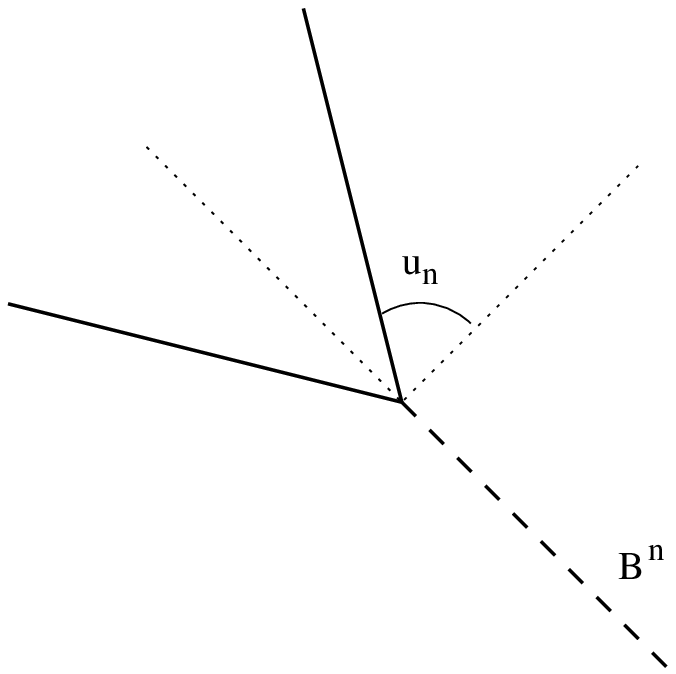}}}\caption{:
Bulk vertices}\label{fig:bulk}~~~~~\end{figure}

In the bulk SG model 
any scattering amplitude among solitons and anti solitons 
factorizes into a product of two particle
scattering amplitudes, of which the independent ones are \cite{ZZ}
\begin{eqnarray*}
a(u)= & S^{++}_{++}(u)=S_{--}^{--}(u)= & -\prod ^{\infty }_{l=1}\left[ \frac{\Gamma (2(l-1)\lambda -\frac{\lambda u}{\pi })\Gamma (2l\lambda +1-\frac{\lambda u}{\pi })}{\Gamma ((2l-1)\lambda -\frac{\lambda u}{\pi })\Gamma ((2l-1)\lambda +1-\frac{\lambda u}{\pi })}/(u\to -u)\right] \\
b(u)= & S^{+-}_{+-}(u)=S_{-+}^{-+}(u)= & 
\frac{\sin (\lambda u)}{\sin (\lambda (\pi -u))}a(u)\quad\qquad;\qquad
\lambda=\frac{8\pi}{\beta^2}-1=\frac{1}{p}\,,\\
c(u)= & S^{-+}_{+-}(u)=S_{-+}^{+-}(u)= & 
\frac{\sin (\lambda \pi )}{\sin (\lambda (\pi -u))}a(u)\quad \qquad ;\qquad u=-i\theta \,\, \, .
\end{eqnarray*}
Since we are concentrating on the bound state poles located at
purely imaginary rapidities we use the variable \( u \) instead of
\( \theta  \) and refer to it as the rapidity from now on in this
Section. The other scattering amplitudes can be described in terms
of the functions 
\[
\{y\}=\frac{\left( \frac{y+1}{2\lambda }\right) \left( \frac{y-1}{2\lambda }\right) }{\left( \frac{y+1}{2\lambda }-1\right) \left( \frac{y-1}{2\lambda }+1\right) }\quad ,\quad (x)=\frac{\sin \left( \frac{u}{2}+\frac{x\pi }{2}\right) }{\sin \left( \frac{u}{2}-\frac{x\pi }{2}\right) }\, \, \, ,\]
as follows. For the scattering of the breathers \( B^{n} \) and \( B^{m} \)
with \( n\geq m \) and relative rapidity \( u \) we have \cite{ZZ}
\[
S^{n\, m}(u)=S^{n\, m}_{n\, m}(u)=\{n+m-1\}\{n+m-3\}\dots \{n-m+3\}\{n-m+1\}\, \, \, ,\]
while for the scattering of the soliton (anti soliton) and \( B^{n} \)
we have

\[
S^{n}(u)=S_{+\, n}^{+\, n}(u)=S_{-\, n}^{-\, n}(u)=\{n-1+\lambda \}\{n-3+\lambda \}\dots \left\{ \begin{array}{c}
\{1+\lambda \}\quad \textrm{if }n\textrm{ is even}\\
-\sqrt{\{\lambda \}}\quad \textrm{if }n\textrm{ is odd}\, \, \, .
\end{array}\right. \]
All the poles of the scattering amplitudes in the physical strip originate
from virtual processes either in the forward or in the cross channel
of the diagrams (a-b) on Fig.(\ref{fig:bulk}), where the useful definition
\[ u_{n}=\frac{n\pi }{2\lambda } \] was also introduced.\footnote{On all the space-time
diagrams time develops from top to bottom. Solitons (or anti solitons)
are denoted by solid lines, while breathers by dashed ones.} For each
such process a coupling as \( f_{n\, m}^{n+m} \) or \( f_{+-}^{n} \)
can be attributed and it is known that \( f_{+-}^{n}=(-1)^{n}f_{-+}^{n} \). 

\subsection{Reflection factors on ground state boundary}

The reflection factor of the soliton on the ground state Neumann boundary
was found by Ghoshal and Zamolodchikov \cite{GZ}. For the topological
charge preserving process, (i.e. when a soliton (anti soliton)
reflects as soliton (anti soliton)), we have \[
P(u)=\cos (\lambda u)R_{0}(u)\sigma \left( \frac{\pi }{2}(\lambda +1),u\right) \sigma (0,u)\, \, \, ,\]
where \[
R_{0}(u)=\prod ^{\infty }_{l=1}\left[ \frac{\Gamma (4l\lambda -\frac{2\lambda u}{\pi })\Gamma (4\lambda (l-1)+1-\frac{2\lambda u}{\pi })}{\Gamma ((4l-3)\lambda -\frac{2\lambda u}{\pi })\Gamma ((4l-1)\lambda +1-\frac{2\lambda u}{\pi })}/(u\to -u)\right] \]
 is the boundary condition independent part and \[
\sigma (x,u)=\frac{\cos x}{\cos (x+\lambda u)}\prod ^{\infty }_{l=1}\left[ \frac{\Gamma (\frac{1}{2}+\frac{x}{\pi }+(2l-1)\lambda -\frac{\lambda u}{\pi })\Gamma (\frac{1}{2}-\frac{x}{\pi }+(2l-1)\lambda -\frac{\lambda u}{\pi })}{\Gamma (\frac{1}{2}-\frac{x}{\pi }+(2l-2)\lambda -\frac{\lambda u}{\pi })\Gamma (\frac{1}{2}+\frac{x}{\pi }+2l\lambda -\frac{\lambda u}{\pi })}/(u\to -u)\right] \]
describes the boundary condition dependence.\footnote{%
The function \( \sigma (x,u) \) has among others the property \( \sigma (x,u)\sigma (x,-u)=\frac{\cos ^{2}x}{\cos (x+\lambda u)\cos (x-\lambda u)} \)
which corrects a typo in \cite{GZ} and \cite{gosh}.
}The poles of \( \sigma (x,u) \) in the physical strip are located
at \( u=\frac{x}{\lambda }-u_{2k+1}\) or at \(\pi-\frac{x}{\lambda
}+u_{2k+1}\), \(k\geq 0 \) ; and it
has no zero there. 

For the topological charge changing process, (when the soliton (anti soliton)
comes back as anti soliton (soliton)), we have\[
Q(u)=\frac{\sin (\lambda u)}{\sin (\frac{\lambda \pi }{2})}P(u)\, \, \, .\]
Note that the topological charge is changed by two in this process,
thus the parity of the soliton number is conserved and we have an odd
and an even sector. 

The breather reflection factors share the same structure as the solitonic
ones as a consequence of the general expression in \cite{gosh}. For
Neumann boundary condition and breather \( B^{n} \) they have the following
form\[
R^{(n)}(u)=R_{0}^{(n)}(u)S^{(n)}\left( \frac{\pi }{2}(\lambda +1),u\right) S^{(n)}(0,u)\]
where \[
R_{0}^{(n)}(u)=\frac{\left( \frac{1}{2}\right) \left( \frac{n}{2\lambda }+1\right) }{\left( \frac{n}{2\lambda }+\frac{3}{2}\right) }\prod ^{n-1}_{l=1}\frac{\left( \frac{l}{2\lambda }\right) \left( \frac{l}{2\lambda }+1\right) }{\left( \frac{l}{2\lambda }+\frac{3}{2}\right) ^{2}}\quad ;\quad S^{(n)}(x,u)=\prod ^{\frac{n-1}{2}}_{l=\frac{1-n}{2}}\frac{\left( \frac{x}{\lambda \pi }-\frac{1}{2}+\frac{l}{\lambda }\right) }{\left( \frac{x}{\lambda \pi }+\frac{1}{2}+\frac{l}{\lambda }\right) }\]
 In general \( R_{0}^{(n)} \) would describe the boundary independent
properties and the other factors would give the boundary dependent
ones. In the Neumann case, however, there are coincidences among the
poles and zeros of the various factors and it is better to rewrite
the reflection factor as\[
R^{(n)}(u)=\langle 2n-1\rangle \langle 2n-3+4\lambda \rangle \dots \langle 2(n-k)-1+(1-(-1)^{k})2\lambda \rangle \dots \left\{ \begin{array}{c}
\langle 1\rangle \quad \textrm{if }n\textrm{ is odd}\\
\langle 1+4\lambda \rangle \quad \textrm{if }n\textrm{ is even}
\end{array}\right. \, \, \, ,\]
where\[
\langle y\rangle =\frac{\left( \frac{y+1}{4\lambda }\right) \left( \frac{y-1}{4\lambda }\right) }{\left( \frac{y+1}{4\lambda }-\frac{1}{2}\right) \left( \frac{y-1}{4\lambda }+\frac{1}{2}\right) }\]
is the {}``half-block{}'' with respect to \( \{y\} \).

\subsubsection{Pole analysis}

Now consider the poles of the soliton reflection factor \( P(u) \). 

\vspace{0.3cm}
{\centering \begin{figure}\subfigure[Soliton bulk
pole]{\resizebox*{!}{4cm}{\includegraphics{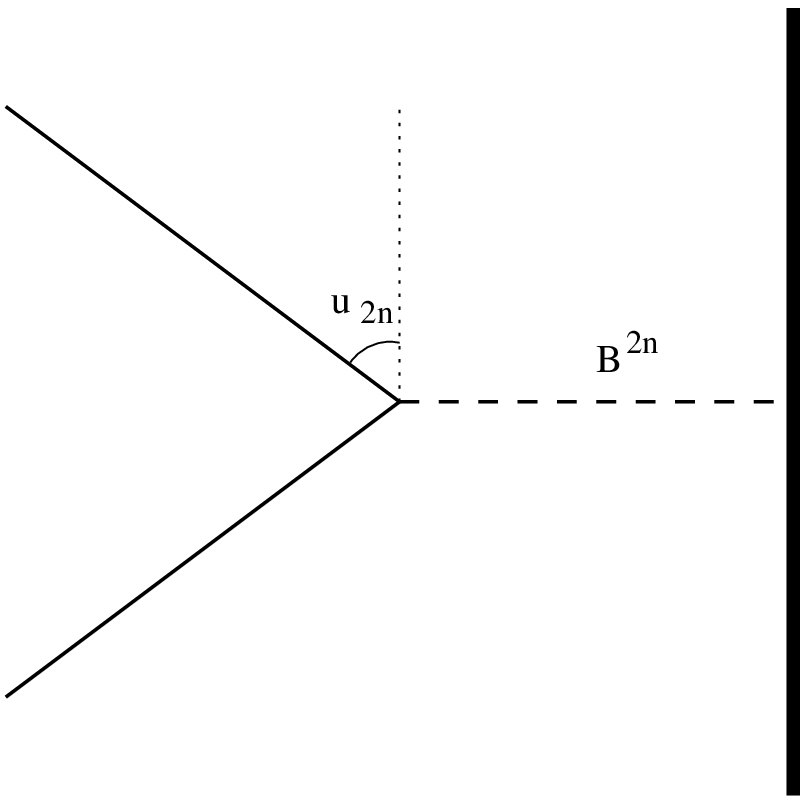}}}
~~~~~~~\subfigure[Breather
triangle]{\resizebox*{!}{4cm}{\includegraphics{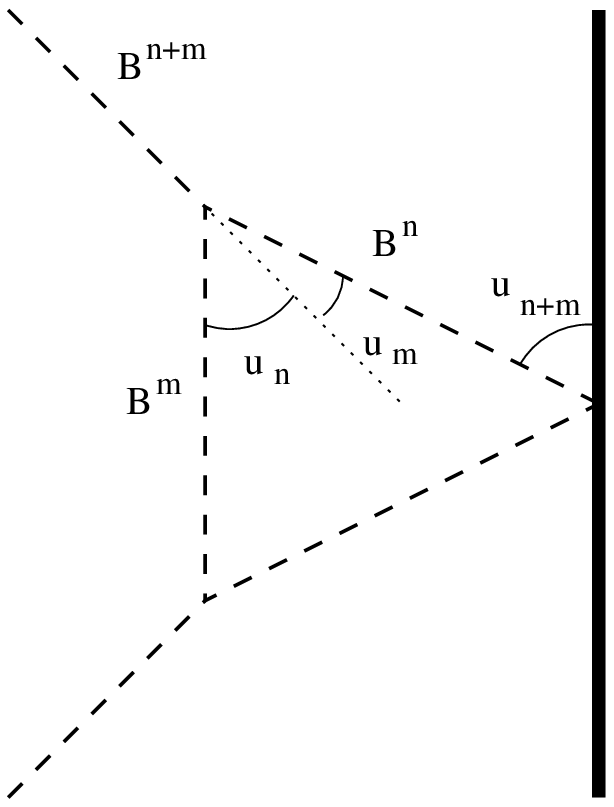}}}
~~~~~~~~~\subfigure[Breather emisson
reflection]{\resizebox*{!}{4cm}{\includegraphics{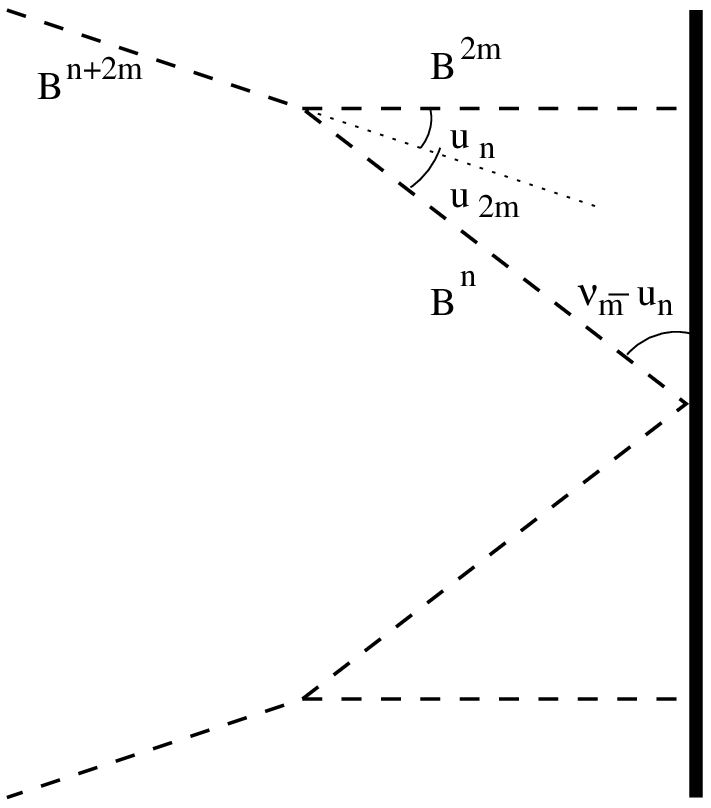}}}
~~~~~~~~\subfigure[bound state
replacement]{\resizebox*{!}{4cm}{\includegraphics{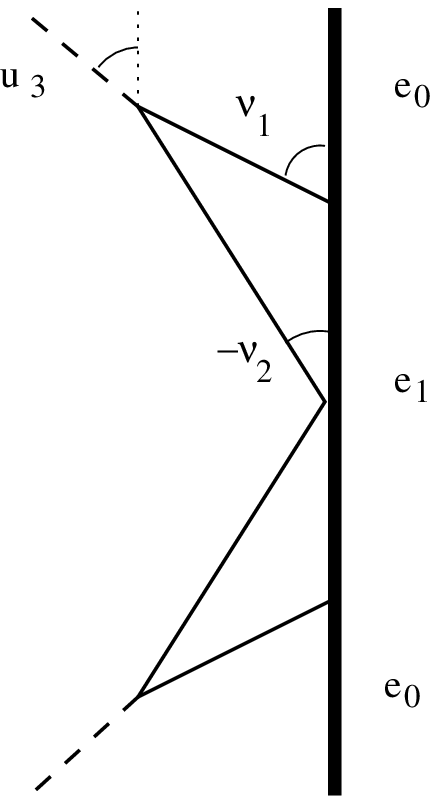}}}\caption{:
A few boundary diagrams}\label{fig:bound} \end{figure}\par}
%\vspace{0.2cm}

\begin{itemize}
\item There are boundary independent poles in the physical strip due to the factor
\( R_{0}(u) \). They are located at \( u_{n}\, ,\, n=1,2,\dots  \)
from which every odd is canceled by the zero of \( \cos (\lambda u) \).
The remaining poles can be described by diagram (a) on
Fig.(\ref{fig:bound}). 
This is consistent with the fact that the \({\bf C }\) symmetric
Neumann boundary can emit and absorb only even breathers with zero
momentum. 
\item The boundary dependent poles in the physical strip are located at\[
\nu _{n}=\frac{\pi }{2}-u_{2n}\quad ;\qquad n=0,1,\dots ,\left[ \frac{\lambda }{2}\right] \]
 The pole at \( \frac{\pi }{2} \) is responsible for the emission
and absorption of a zero momentum soliton or anti soliton. For each
of the other poles we associate a boundary bound state with energy
\[
e_{n}-e_{0}=M\sin u_{2n}\]
 From now on  we put the boundary ground state energy
to zero \( e_{0}=0 \) for simplicity. 
\end{itemize}
Clearly the basically bulk process (a) on Fig.(\ref{fig:bound})  does 
not exist for the reflection
amplitude \( Q(u) \) and the factor \( \sin (\lambda u) \) takes
care of this. The boundary dependent poles of \( Q(u) \) coincide
with and can be explained in the same way as those of \( P(u) \). 

Now let us focus on the poles of the breather reflection factors.
As a warm-up example consider the reflection factor of the third breather:\[
R^{(3)}(u)=\langle 5\rangle \langle 3+4\lambda \rangle \langle 1\rangle \]

\begin{itemize}
\item It has a simple bulk pole at \( \frac{\pi }{2}-u_{3} \) whose explanation
depends on the value of the parameter \( \lambda  \). If \( B^{6} \)
is in the spectrum then we can use the {}\lq all-breather{}' version
of diagram (a). On this we mean by replacing each soliton line by
a \( B^{3} \) line having rapidity \( \frac{\pi }{2}-u_{3} \) and
\( B^{6} \) is formed at the vertex. Now if \( B^{6} \) is not in
the spectrum then the soliton version of diagram (b) applies.
In this case the breather triangle is turned into a soliton-anti soliton
triangle. 
\item The pole at \( \frac{\pi }{2}-u_{1} \) is of second order and can
be explained by diagram (c).
\item The pole at \( u_{3} \) can be explained by the soliton version
of 
diagram (c) for \( 3>\left[ \frac{\lambda }{2}\right]  \) (which is
necessary to guarantee that the soliton in the middle also travels
towards the wall). This diagram gives a second order pole and the reflection
factor has no zero at \( -\nu _{3}=u_{6}-\frac{\pi }{2} \). We have
an analogous diagram, however, by changing the amplitude \( P(u) \)
for \( Q(u) \). They differ by a factor of 
\( \sin (-\lambda \nu _{3})/\sin (\frac{\lambda \pi }{2})=-(-1)^{3} \)
but we know that \( f_{+-}^{3}=(-1)^{3}f_{-+}^{3} \). This gives
the sign difference which is responsible for the Coleman-Thun type
cancellation in the sum of the two diagrams. In the \( 3\leq \left[ \frac{\lambda }{2}\right]  \)
case we have a boundary bound state with energy\[
e_{3}=M_{3}\cos u_{3}=2M\sin u_{3}\cos u_{3}=M\sin u_{6}\ ,\]
 which is exactly the same as the boundary bound state created by
the soliton. The identity of these two boundary bound states can also
be confirmed by showing the agreement of the other higher spin conserved
charges. 
\item For the other two poles at \( u_{1} \) and \( u_{2} \) we have diagram
(b). In the case of  \( u_{2} \)  the second breather hits the boundary
while in the  case of \( u_{1} \)  the first. 
\end{itemize}
We could be satisfied now and turn to the general case. There is a
subtlety, however, since we have not computed the residues. Performing
the computation reveals that the residues of the reflection factors
and diagrams agree except for the pole \( u_{1} \) where they have
different signs. Thus we must assume that a boundary bound state exists
with energy \( M_{3}\cos u_{1}=e_{1}+e_{2} \). We note that in the
Neumann case the bulk pole of \( R^{(3)}_{0}(u) \) and a boundary
pole of \( S^{(3)}(\frac{\pi }{2}(\lambda +1),u) \) at \( u_{1} \)
coincide but a zero in the boundary factor reduces the singularity
to first order. If we move a bit away from the Neumann point, when
the boundary dependent part becomes \( S^{(3)}(\frac{\pi }{2}(\lambda +1)-\delta \lambda ,u) \),
the two poles split and the bulk pole at \( u_{1} \) can be described
by the diagram above, while the pole at \( u_{1}-\delta  \) corresponds
to a boundary bound state. Computing the residue and taking the \( \delta \to 0 \)
limit do not commute, however, resulting in a sign difference. This
sign difference persists when the state with energy \( e_{2} \) is
not in the spectrum i.e. for \( u_{4}>\frac{\pi }{2} \). In this
case we have diagram (d), where \( B^{3} \) is decaying into a soliton
and an anti soliton. Then the soliton creates the state with energy
\( e_{1} \) on which the other anti soliton is reflecting with rapidity
\( u_{4}-\frac{\pi }{2} \). This diagram is naively second order,
however, the same Coleman-Thun mechanism that we used in the previous
example, takes care of the cancellation.

From this example we can draw two conclusions. First, breathers and
solitons can create the same states. Second, even if a diagram describes
a pole of the same order as the reflection factor has, the residues do not
always coincide, thus allowing the creation of a boundary bound state
in the same time. 

Now we are ready to discuss the general case. The reflection factor
\( R^{(n)}(u) \) has the following poles in the physical strip. 

\begin{itemize}
\item There is a simple pole at \( \frac{\pi }{2}-u_{n} \). It can be explained
in terms of the breather version of diagram (a) by forming 
\( B^{2n} \) or if it is not in the spectrum then by the soliton
version of diagram (b). 
\item The double poles at \( \frac{\pi }{2}-u_{n-2k} \) can be explained
by diagram (c). 
\item For even breathers there is an extra simple pole at \( \frac{\pi }{2} \).
It corresponds to the emission of a zero momentum even breather. 
\item The simple pole at \( u_{n} \) corresponds to the creation of the
boundary bound state with energy \( e_{n}=M\sin u_{2n} \), if it
is in the spectrum. Alternatively, it can be explained in terms of
the solitonic version of diagram (c) where, just as in the example,
a Coleman-Thun type mechanism ensures the pole to be simple. 
\item If \( n+k \) is even then the pole \( u_{k} \) is responsible for
the creation of a boundary bound state with energy \( e_{\frac{n+k}{2}}+e_{\frac{n-k}{2}} \),
if \( e_{\frac{n+k}{2}} \) is in the spectrum. Alternatively, we
have diagram (d) where the state with energy \( e_{k} \) is created
in the middle and the Coleman-Thun argument has to be used for the
scattering. 
\item If, however, \( n+k \) is odd then the pole at \( u_{k} \) has its
explanation in terms of diagram (b). 
\end{itemize}

\subsection{Excited state reflection factors }

We have seen that both the pole of the soliton reflection factor at \( \nu _{n} \)
and the pole of the reflection factor of \( B^{n} \) at \( u_{n} \)
give rise to a state with energy \( e_{n} \). The identity of these
two 
states can be further confirmed by showing that any reflection factors
on them agree. Computing these reflection factors from the solitonic
and breather bootstrap equations gives identical results. Therefore    
 we denote these common states by \( |n\rangle  \). 

In computing the soliton excited state reflection factor we use the
\[
P_{|n\rangle }(u)=P(u)S^{n}(u+u_{n})S^{n}(u-u_{n})\]
{}''soliton-breather{}'' bootstrap equation. The result is \[
P_{|n\rangle }(u)=a_{n}(u)P(u)\quad ;\qquad a_{n}(u)=\{2n-1+\lambda \}\{2n-3+\lambda \}\dots \{1+\lambda \}\, \, \, .\]
We have checked that the purely solitonic bootstrap equation gives
the same result. Since the solitonic reflection factor has one extra
pole at \( \nu _{0}=\frac{\pi }{2} \) we can bootstrap on 
this pole and obtain \( P_{|0\rangle }(u)=P(u) \). This means that
the ground state is non degenerate, so the creation or annihilation of
a soliton or anti soliton does not change the boundary ground state.
(We have already used  this property in the explanation of the ground
state breather poles). The existence of a unique ground state implies
that ${\bf C}$ symmetry is not broken.  
Note, however, that this is strikingly different
from the case for Dirichlet boundary condition at \( \xi =\frac{\pi }{2}(\lambda +1) \),
where one has two vacua and the \( Z_{2} \) symmetry is spontaneously
broken \cite{saj}.  

Repeating the same bootstrap program for the amplitude \( Q(u) \)
we arrive at\[
Q_{|n\rangle }(u)=a_{n}(u)Q(u)\, \, \, .\]
 For the breather reflection factors on the excited state \( |n\rangle  \)
the bootstrap equation takes the form
\[
R^{(m)}_{|n\rangle }(u)=R^{(m)}(u)S^{n\, m}(u+u_{n})S^{n\, m}(u-u_{n})\, \, \, .\]
From here we have explicitly \[
R^{(m)}_{|n\rangle }(u)=b^{m}_{n}(u)R^{(m)}(u)\]
where
\[\eqalign{
&b^{m}_{n}(u)=\{m+2n-1\}\{m+2n-3\}\dots \cr &\dots\{m+2n-1-2l\}\dots \left\{ \begin{array}{c}
\{m-2n+1\}\quad \textrm{if}\quad m-2n+1>0\\
\{2n-m+1\}\quad \textrm{if}\quad 2n-m+1>0
\end{array}\right. \, \, \, .}\]

Let us focus on the solitonic reflection factors $P_{\vert n\rangle }(u)$. 
The prefactor \( a_{n}(u) \)
has simple poles at \( \nu _{0} \) and \( \nu _{n} \) and 
double poles at \( \nu _{k}\, ,\, \, k=1,2,\dots ,n-1 \). The
locations 
of these poles are the same as that of \( P(u) \), thus only the
order has increased for \( k\leq n \). We conjecture, and in fact
prove in Appendix A, that the other simple poles at \( \nu _{k}\, ,\, \, k>n \),
which do not come from \( a_{n}(u) \), are responsible for the creation
of the excited boundary bound states with energy \( e_{n}+e_{k} \)
which we denote by \( |n,k\rangle  \). 

Now it is interesting to use the integrability of the model. Using
the factorizability argument we can shift the soliton trajectory and
obtain the three diagrams on Fig.(\ref{fig:ujall}). 

{\centering
\begin{figure}~~~~\subfigure{\resizebox*{!}{5cm}{\includegraphics{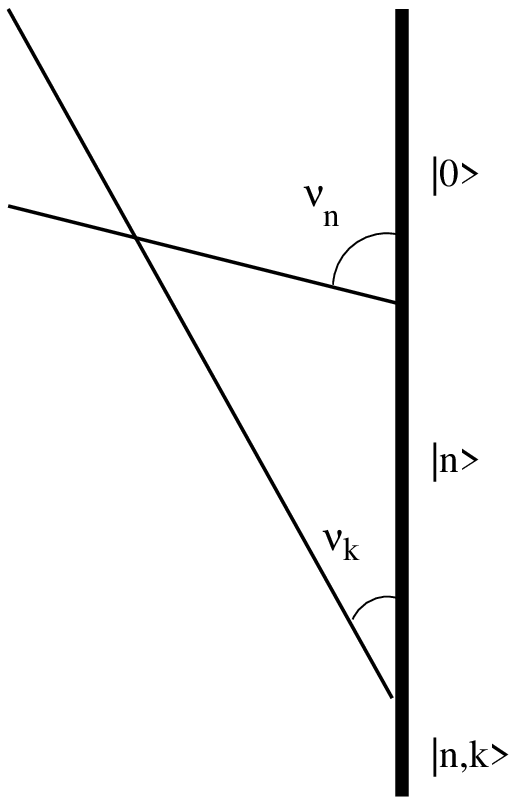}}}
~~~~~~~~~~~~~~~~\subfigure{\resizebox*{!}{5cm}{\includegraphics{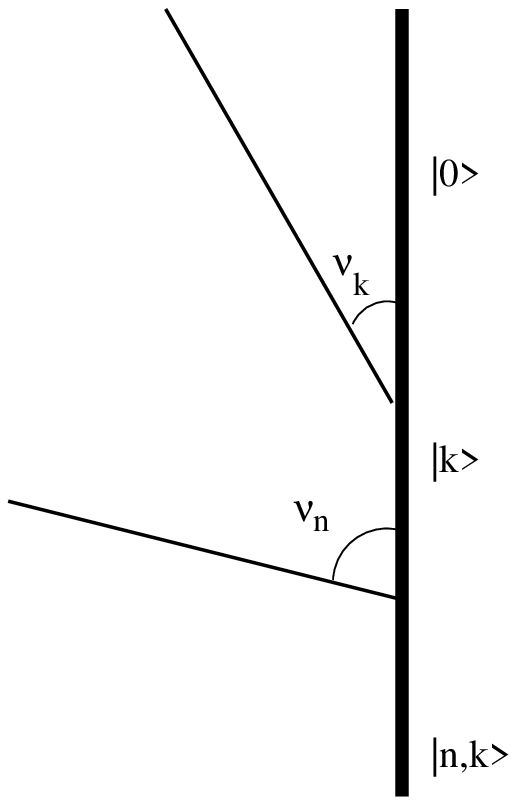}}}
~~~~~~~~~~~~~~~~~~\subfigure{\resizebox*{!}{5cm}{\includegraphics{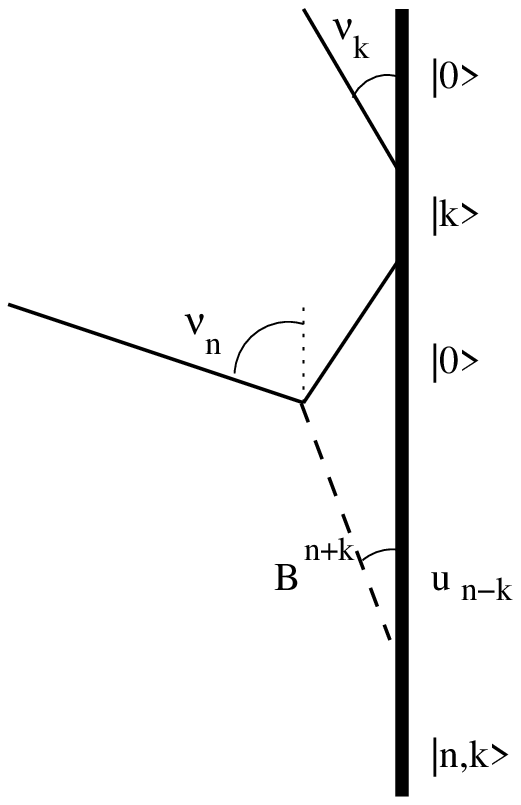}}}
\caption{: Various ways of creating the state $\vert n,k\rangle $.}
\label{fig:ujall}\end{figure}\par}
%\vspace{0.3cm}

In the first picture the soliton creates the boundary bound states
\( |n\rangle  \) at an angle \( \nu _{n} \) then an other soliton creates
\( |n,k\rangle  \) at an angle \( \nu _{k} \) where \( k>n \). They scatter
on each other at an angle  \( u_{2(k-n)} \) which is not proper for forming
a breather. Now if we move the second soliton trajectory up we obtain
the picture in the middle. Here the soliton at \( \nu _{k} \) creates
the state \( |k\rangle  \) then the other soliton reflects on it
at an angle \( \nu _{n} \). This reflection factor has a third order pole
for which diagram (a) on Fig.(\ref{fig:1sol2br}) shown in Appendix B
can be 
associated showing the creation of the state \( |n,k\rangle  \)
as it was expected from consistency. In the third picture the excited
boundary \( |k\rangle  \) decays by emitting a soliton. This soliton
then scatters on the other soliton forming \( B^{n+k} \) with rapidity
\( u_{k-n} \) which finally creates the state \( |n,k\rangle  \).  

Now we can use any of the processes above (or their breather versions)
to compute the reflection amplitudes. They turn out to be\[
P_{|n,k\rangle }(u)=a_{n}(u)a_{k}(u)P(u)\quad ;\quad Q_{|n,k\rangle }(u)=a_{n}(u)a_{k}(u)Q(u)\]
and \[
R^{(m)}_{|n,k\rangle }(u)=b^{m}_{n}(u)b^{m}_{k}(u)R^{(m)}(u)\]
From this it is clear how it goes on, so we turn to the description
of the spectrum of the boundary bound states and their reflection
factors in general.

\subsection{The spectrum of the boundary bound states and the associated reflection
factors}

From the previous discussion it is clear that a boundary bound state
can be labelled by a sequence of increasing positive integers and
denoted by 
\( |n_{1},n_{2},\dots ,n_{k}\rangle  \) where \( n_{i}>n_{i-1} \)and
\( n_{k}<\left[ \frac{\lambda }{2}\right]  \). Its energy can be
written as \[
m_{|n_{1},n_{2},\dots ,n_{k}\rangle }=\sum ^{k}_{j=1}M\sin u_{2n_{j}}=\sum _{j=1}^{k}e_{n_j}.\]
 The first task is to prove the existence of these states. Unfortunately
the proof of \cite{patr}, which uses the breathers to create the
boundary bound states, does not apply. This proof is based on a
special property of the breather reflection factors with Dirichlet
b.c.-s, namely that to every pole either a bound state or a 
Coleman-Thun diagram can be associated. In contrast to the Dirichlet case we
have shown that 
  Coleman-Thun
diagrams and boundary bound state creations may coexist in the Neumann
problem. We prove, however,
in Appendix A that solitons at reflection angle \( \nu _{m} \)
on the state \( |n_{1},n_{2},\dots ,n_{k}\rangle  \) can create the
state \( |n_{1},n_{2},\dots ,n_{k},m\rangle  \) if \( m>n_{k} \).
Now the next step is to compute the reflection factors and explain
all the poles in terms of diagrams. The computation is quite straightforward
and the result is \begin{eqnarray*}
P_{|n_{1},n_{2},\dots ,n_{k}\rangle }(u) & = & a_{n_{1}}(u)a_{n_{2}}(u)\dots a_{n_{k}}(u)P(u)\\
Q_{|n_{1},n_{2},\dots ,n_{k}\rangle }(u) & = & a_{n_{1}}(u)a_{n_{2}}(u)\dots a_{n_{k}}(u)Q(u)
\end{eqnarray*}
and \[
R^{(m)}_{|n_{1},n_{2},\dots ,n_{k}\rangle }(u)=b^{m}_{n_{1}}(u)b^{m}_{n_{2}}(u)\dots b^{m}_{n_{k}}(u)R^{(m)}(u).\]
Using the factorizability argument leads to the following creation-annihilation
rules: For the soliton with rapidity \( \nu _{m} \)

\[
|n_{1},\dots ,n_{i},n_{i+1},\dots ,n_{k}\rangle \longrightarrow |n_{1},\dots ,n_{i},m,n_{i+1},\dots ,n_{k}\rangle \]
if \( n_{i}<m<n_{i+1} \). For \( B^{m} \) with rapidity \( u_{2n_{i}+m} \)
we have\[
|n_{1},\dots ,n_{i},\dots ,n_{k}\rangle \longrightarrow |n_{1},\dots ,n_{i}+m,\dots ,n_{k}\rangle \, \, \, ,\]
while for \( B^{m+l} \) with rapidity \( u_{l-m} \) and \( l>m \)
the process is \[
|n_{1},\dots ,n_{i},n_{i+1},\dots ,n_{j},n_{j+1},\dots ,n_{k}\rangle \longrightarrow |n_{1},\dots ,n_{i},m,n_{i+1},\dots ,n_{j},l,n_{j+1},\dots ,n_{k}\rangle \, \, \, .\]
Now the most difficult problem is to explain all the poles of the
reflection factors in terms of diagrams. Since it is quite cumbersome
we put it in Appendix B.

\section{Finite size effects in large volume}

\subsection{Bethe-Yang equations for particles in finite volume with boundaries}

Imagine putting \( N \) particles in finite volume \( L \), such that the
\( i \)th particle \( A_{i} \) has mass \( M_{i} \) and rapidity \( \theta _{i} \)
(which we choose nonnegative since only the absolute value matters, due to reflection
on the boundaries), plus some additional internal quantum numbers \( \alpha _{i} \)
(distinguishing it within a multiplet of the same mass \( M_{i} \)). Let us
denote the reflection factor of the \( i \)th particle on the left/right end
of the interval by \( R^{(i)}_{L}(\theta _{i}) \) and \( R^{(i)}_{R}(\theta _{i}) \),
respectively and the scattering matrix of particles \( A_{i} \) and \( A_{j} \)
by \( S^{(i,j)}(\theta _{i}-\theta _{j}) \). Obviously in the general
case these are matrices with the following structure
\begin{eqnarray*}
R^{(i)}_{L}(\theta _{i}) & = & \left\{ R^{(i)}_{L}(\theta _{i})^{\alpha ^{'}_{i}}_{\alpha _{i}}\right\}\,,\\
R^{(i)}_{R}(\theta _{i}) & = & \left\{ R^{(i)}_{R}(\theta _{i})^{\alpha ^{'}_{i}}_{\alpha _{i}}\right\}\,,\\
S^{(i,j)}(\theta _{i}-\theta _{j}) & = & \left\{ S^{(i,j)}(\theta
_{i}-\theta _{j})_{\alpha _{i}\alpha _{j}}^{\alpha _{i}^{'}\alpha
_{j}^{'}}\right\}\,. 
\end{eqnarray*}
We define a family of transfer matrices that acts in \( V_{1}\otimes \dots \otimes V_{N} \),
where \( V_{i} \) is the internal space labelled by the multiplicity index \( \alpha _{i} \)
of the \( i \)th particle. The transfer matrix \( T_{k} \) describes a
unitary transformation on the wave function 
 resulting from the scattering of \( A_{k} \) on the other particles and
the two boundaries when we take the particle around the volume (starting e.g
to the left, reflecting on the left boundary, coming back to the right, reflecting
on the right boundary and returning to the original position):
\[
T_{k}^{(1,\dots ,\, N)}\left( \theta _{1},\dots ,\theta _{N}\right)
 =R^{(k)}_{R}(\theta _{k})\prod _{j:j\neq k}S^{(j,k)}(\theta
 _{k}+\theta _{j})R_{L}^{(k)}(\theta _{k})\prod _{j:j\neq
 k}S^{(j,k)}(\theta _{k}-\theta _{j})\ .\]
 We omitted the multiplicity indices of the matrices (these are indicated by
the upper indices in brackets and can be restored easily). The scatterings were
performed in a particular order, due to the equations of factorized
bulk and boundary scatterings, the order is eventually immaterial: performing
the scatterings in any other way results in the same transfer matrix. Another
consequence of factorized scattering is that the matrices \( T_{k} \) form
a commuting family for a given set of rapidities. The wave function can be characterized
by a vector \( \psi _{\alpha _{1}\dots \alpha _{N}} \) living in the tensor
product space, and transforming under all such monodromy operations as:
\begin{equation}
\label{bethe-yang}
\exp \left( 2iM_kL\sinh (\theta _{k})\right) T^{(1,\dots
,N)}_{k}\left( \theta _{1},\dots ,\theta _{N}\right) \psi =\psi \quad
,\quad k=1,\dots ,N\ .
\end{equation}
The prefactor in this equation is the phase acquired by \( A_{k} \) due to
its momentum (whose absolute value is \( M_k\sinh \theta _{k} \)). Equations
(\ref{bethe-yang}) are the so-called Bethe-Yang equations in the presence of
a boundary.

Since the transfer matrices commute, they can be diagonalized simultaneously.
Let us denote the eigenvalues of the transfer matrix \( T^{(1,\dots ,N)}_{k}\left( \theta _{1},\dots ,\theta _{N}\right)  \)
by \( \lambda ^{(s)}_{k}\left( \theta _{1},\dots ,\theta _{N}\right)  \),
where \( s=1,\dots ,D_{N} \) enumerates the eigenvalues (with multiplicities)
and \( D_{N}=\dim V_{1}\otimes \dots \otimes V_{N} \). The corresponding common
eigenvectors will be called \( \psi ^{(s)}\left( \theta _{1},\dots ,\theta _{N}\right)  \). 

The solutions to the Bethe-Yang equations (\ref{bethe-yang}) are given by the
wave function amplitudes
\[
\psi =\psi ^{(s)}\left( \theta _{1},\dots ,\theta _{N}\right) \]
 provided that the rapidities solve the equations\footnote{For scalar
 particles, these equations appeared in \cite{FendlSal}.} 
\begin{equation}
\label{quant_rel}
\exp \left( 2iM_kL\sinh (\theta _{k})\right) \lambda ^{(s)}_{k}\left( \theta _{1},\dots ,\theta _{N}\right) =1\quad ,\quad k=1,\dots ,N
\end{equation}
for some (fixed) \( s \). These equations effectively describe the quantization
of momentum in finite volume with boundaries. The total kinetic energy of the particles is
then given by 
\[
E_{\mathrm{kin}}=\sum ^{N}_{j=1}M_{j}\cosh \theta _{j}\ ,\]
which is the energy difference with respect to the state with no particles in
this approximation. 

The above calculation supposes that the particles do not overlap with each other
and the boundary substantially, i.e. it is valid only when the volume is large
enough: \( M_{i}L\gg 1 \) for all \( i \). Note that if the particles involved
in the calculation have no multiplicity labels, the transfer matrix is a scalar
phase itself and can be directly substituted into Eqs. (\ref{quant_rel}).
Taking the logarithm of Eqs. (\ref{quant_rel}) yields
\begin{equation}
\label{log_quant_rel}
2M_kL\sinh (\theta _{k})-i\log \lambda ^{(s)}_{k}\left( \theta
_{1},\dots ,\theta _{N}\right) =2\pi I_{k}\quad ,\quad k=1,\dots ,N\ .
\end{equation}
where the \( I_{k} \) are integers that we call Bethe quantum numbers. In this
way a given multi-particle state can be labeled by giving \( s \) and a set
of integers \( I_{1}\, ,\dots \, ,I_{N} \) and this labeling is normally unique
given a consistent choice of the branch of the logarithm. There are some exceptions
when for given \( s \) and \( I_{1}\, ,\dots \, ,I_{N} \) Eqs. (\ref{log_quant_rel})
admit more than one solution; such cases will be explicitly indicated below.

Note that in these considerations we neglected the contribution of
 vacuum polarization. Since, on dimensional grounds, we expect this contribution to 
behave as ${\cal O}(\exp (-ML))$ for large $L$-s (with $M$ being some characteristic 
mass scale), the polynomially decreasing, leading finite size
 corrections are indeed 
described by 
Eq.(\ref{log_quant_rel}).
     
\subsection{Some particular cases}

Here we write down and analyze Eqs. (\ref{log_quant_rel}) for some simple
configurations that will play an important role in the sequel.

\subsubsection{States containing a single scalar particle}

For a state containing a single scalar particle of mass \( M \), Eqs. (\ref{log_quant_rel})
reduce to the single equation
\begin{equation}
\label{scalar_quant_cond}
2ML\sinh (\theta )-i\log R_{L}\left( \theta \right) -i\log R_{R}\left(
\theta \right) =2\pi I\ .
\end{equation}
The energy of the state with respect to the vacuum can be written as
\[
E(L)-E_{0}(L)=E_{R}+E_{L}+M\cosh \theta\ , \]
where \( E_{R,L} \) denote the boundary energy of the left and right boundaries
with respect to the boundary ground state (note that the boundary can also be
in an excited state). One can easily express the function \( E(L)-E_{0}(L) \)
in a parametric form using \( \theta  \).

Our studies show that for \( ML\ll 1 \) Eqn. (\ref{scalar_quant_cond}) always
has a real solution irrespective of the choice of the quantum number \( I \).
However this may not be true for larger values of  \( ML \). Let us discuss the particular
case \( I=0 \) and introduce the notation
\[
\rho _{L,R}=\left. i\frac{\partial }{\partial \theta }
\log R_{L,R}(\theta )\right| _{\theta =0}\,.\]
It is a generic property of reflection factors that \( i\log R_{L,R}(\theta ) \)
has a finite limit as \( \theta \, \rightarrow \, \pm \infty  \); it also
follow from unitarity that they are odd functions. Therefore the equation
\begin{equation}
\label{I_0_quant_rel}
2ML\sinh (\theta )=i\log R_{L}\left( \theta \right) +i\log R_{R}\left( \theta \right) 
\end{equation}
has a pair of nonzero solutions as long as 
\[
2ML<\rho _{L}+\rho _{R}\ .\]
Increasing \( L \), these two solutions move towards the origin, which they
reach at a finite volume 
\[
L_{0}=\frac{1}{2M}\left( \rho _{L}+\rho _{R}\right)\ . \]
The question arises: what happens for volumes \( L>L_{0} \)? If \( ML_{0}\gg 1 \)
the Bethe-Yang equations should give a good approximation of the finite volume
spectrum around this point. However, eigenvalues cannot simply disappear as
the Hamiltonian we are concerned with is Hermitian. In fact, the spectrum is
real and every eigenvalue is a continuous (even smooth) function of \( L \).
This leads us to the description of boundary bound states.

\subsubsection{Description of boundary bound states in finite volume}

Quite naturally, the two solutions of Eqn. (\ref{I_0_quant_rel}) do not disappear,
but continue their life as complex (eventually imaginary) solutions. Substituting
\( \theta =iu \) into Eqn. (\ref{I_0_quant_rel}) we obtain
\begin{equation}
\label{imquantrel}
2ML\sin (u)=\log R_{L}\left( iu\right) +\log R_{R}\left( iu\right)\ . 
\end{equation}
As reflection factors always take real values on the imaginary axis (as a result
of analytic \( S \)-matrix theory) this is a real equation and it has a pair
of real solutions for \( u \) exactly when \( L>L_{0} \). The energy corresponding
to this solution is
\[
E(L)-E_{0}(L)=E_{R}+E_{L}+M\cos u\ .\]
This is smaller than \( E_{R}+E_{L}+M \) which means that this state cannot
correspond to a real particle between the two boundaries. 

The reflection factors \( R_{L} \) and \( R_{R} \) normally have poles on
the imaginary axis. Let us denote the one which is closest to the origin by
\( u^{*} \) and suppose it occurs in \( R_{L} \) only\footnote{We meet 
this situation e.g. in sine-Gordon model with $\phi\vert_R=0$ 
(special Dirichlet) and $\partial_x\phi\vert_L=0$ (Neumann) b.c-s.}.   
Then one has
\[
u\, \rightarrow \, u^{*}\quad \mathrm{as}\quad L\, \rightarrow \, \infty \]
and so
\[
E(L=\infty )-E_{0}(L=\infty )=E_{R}+E_{L}+M\cos u^{*}\ .\]
Suppose also that \( u^{*} \) corresponds to a boundary bound state
$\vert B_L^{*}\rangle $ 
in the bootstrap, then the energy of such a state is
\[
E_{L^{*}}=E_{L}+M\cos u^{*}\ .\]
It is then tempting to interpret the corresponding state as one without particles
but in which the left boundary is excited to the state corresponding to the
pole at \( u=u^{*} \). Indeed our numerical data show complete consistency
with this interpretation. 

When the left and the right boundary conditions  
 are identical (e.g both are Neumann), both reflection factors
have a pole at \( u=u^{*} \). One can then interpret the resulting 
\lq finite size state' as
one in which one of the boundaries is an excited one. This can be realized with
two wave functions (as we have two identical boundaries) and we expect that
the state described above corresponds to one of them. Note that in the above
solution we always have \( u<u^{*} \). The other one can be described (at least
for large enough \( L \)) by noting that there is going to be another solution
\( u' \) to Eqn. (\ref{imquantrel}) which satisfies \( u'>u^{*} \) and also
approaches \( u^{*} \) for \( L\, \rightarrow \, \infty  \). It is clear, 
that, for finite $L$, the solution with $u'$ has a lower energy than that of
with $u$. 

When $L$ and $R$ are not identical 
and the boundary state, $\vert B_R\rangle $, is not identical to the excited 
$\vert B_L^{*}\rangle $ either, one again expects two wave functions 
for the system with no particles, so the two solutions that exist in 
this case too can
be interpreted once again in this way. However, when 
$\vert B_R\rangle =\vert B_L^{*}\rangle $ only one
of the solutions can have an interpretation in terms of a real physical state.
This shows that not necessarily all solutions of the Bethe-Yang 
equations (\ref{bethe-yang})
correspond to physical states. In fact, when \( u^{*} \) corresponds to a Coleman-Thun
pole, we do not even expect the \( I=0 \) state to appear in the spectrum,
since then we would have some state for \( L<L_{0} \) that does not have any
physical continuation to \( L_{0}<L \).

There is a generalization of this line of thought to other poles that are 
further
away from the origin in the physical strip \( 0<u<\pi /2 \) . We cannot expect
these to be realized in an analytic continuation from a state with real rapidities,
since any such continuation goes through the origin \( u=\theta =0 \) and
necessarily runs into the pole closest to \( 0 \). However, we can calculate
solutions of Eqn. (\ref{imquantrel}) around any such pole (of course one cannot
go too far from the given pole since one then generally runs into 
the domain of some other
singularity of the reflection factor). For large enough \( L \), such a solution
exists and generalizing the above scheme one expects that it can describe the
leading finite size behaviour of some excited boundary state in finite volume
if the pole in question participates in the bootstrap. Once again, Coleman-Thun
poles can have no such corresponding state in the finite size spectrum. 

A given excited boundary state may arise from different reflection factors as
a pole (e.g. in SGN the state $\vert n\rangle $ arises from both the
soliton and the $B^n$ reflections) 
and therefore a further possibility is that it can be described in finite
volume in more than one ways. 

Later in this paper we show that all the above considerations are well confirmed
by our numerical studies. We wish to remark that similar ideas to
describe bound states appeared in \cite{ktw} for the bulk and in
\cite{dptw} for the boundary case.

\subsubsection{States with two scalar particles}

In the case of two scalar particles of mass \( M_{1} \) and \( M_{2} \), the
Bethe-Yang equations take the form
\be\label{2egy}\eqalign{
2M_{1}L\sinh \theta _{1}-i\log R^{(1)}_{L}(\theta _{1})-&i\log
R^{(1)}_{R}(\theta _{1})-i\log S^{(12)}(\theta _{1}-\theta _{2})\cr &-i\log
S^{(12)}(\theta _{1}+\theta _{2}) =2\pi I_{1},\cr 
2M_{2}L\sinh \theta _{2}-i\log R^{(2)}_{L}(\theta _{2})-&i\log
R^{(2)}_{R}(\theta _{2})-i\log S^{(12)}(\theta _{2}-\theta _{1})\cr &-i\log
S^{(12)}(\theta _{2}+\theta _{1})=2\pi I_{2}},
\ee
\[
E(L)-E_{0}(L)=E_{R}+E_{L}+M_{1}\cosh \theta _{1}+M_{2}\cosh \theta
_{2}\ .\]
In this case, these equations can only be solved by numerical iteration. One
can also continue these equations to imaginary value of one (or both) of the
rapidities. Using that the reflection factors are real for
purely imaginary rapidities together with unitarity and real
analyticity of the bulk $S$ matrix  one can show 
(just as for the one-particle case) that the resulting
equations are consistent and the energy of the solutions is always
real. Consider e.g. the Bethe Yang equations for $B^n$-$B^m$
with rapidities $iu$ and $\theta_2$ respectively, which, 
before taking the logarithm, can be written as
\be\label{2egyv}
\eqalign{
R_L^{(n)}(iu) R_R^{(n)}(iu)\vert S^{(n,m)}(\theta_2+iu)\vert^2{\rm
e}^{-2m_nl\sin u}=&1\,,\qquad l=ML,\cr
R_L^{(m)}(\theta_2) R_R^{(m)}(\theta_2)\frac{S^{(n,m)}(\theta_2+iu)}{S^{(n,m)}(\theta_2+iu)^*} {\rm
e}^{i2m_ml\sinh\theta_2}=&1\,,\qquad m_{n,m}=\frac{M_{n,m}}{M}.}
\ee
The first is a real equation with real entries, and each factor on
the left hand side of the second equation is of unit modulus. Thus,
when taking the logarithms, for the complete system there is only one
quantum number, $I_m$, coming from the second equation.    
Two particle states with one rapidity being
imaginary are important, as one 
 can argue that the best way to 
describe one  particle states moving between an excited and a
ground state boundaries is to use (\ref{2egy}-\ref{2egyv})  with {\sl ground state}
reflection factors and an imaginary rapidity tuned into the vicinity
of a pole corresponding to the excited state. When both rapidities are
imaginary, the solutions of (\ref{2egy}) describe zero particle states
with excited boundaries.

\section{Truncated Conformal Space Approach (TCSA) for the boundary
sine-Gordon model}

Here we describe the Hamiltonian of boundary sine-Gordon
model (BSG) living on the line segment $0\le x\le L$ as that of a bulk
 and boundary perturbed free boson 
with suitable boundary conditions.
This is the starting point of the TCSA analysis. 

%\subsection{Bulk and boundary perturbations}

The basic idea of TCSA is to describe certain $2d$ models as relevant
perturbations of their ultraviolet limiting CFT-s \cite{YZ}. If we consider
boundary field theories, then the CFT-s in the ultraviolet are in fact
boundary CFT-s. The use of TCSA to investigate boundary theories was
advocated in \cite{dptw,ger}.

As the bulk SG can be successfully described as a perturbation of the
$c=1$ free boson \cite{Gab}, it is natural to expect, that the various BSG models
are appropriate perturbations of $c=1$ theories with Neumann or
Dirichlet boundary conditions. Therefore we  
take the strip $0\le x\le L$ and consider the following perturbations
of the models described in detail in Appendix C:
\begin{eqnarray*}
S & = & \int ^{\infty }_{-\infty }\int _{0}^{L}\left( \frac{1}{8\pi }\partial _{\mu }\Phi \partial ^{\mu }\Phi +\mu \cos (\beta \Phi )\right) dxdt+\\
 &  & +\int ^{\infty }_{-\infty }\left( M_{0}\left[\cos \left( \frac{\beta
}{2}(\Phi _{B}-\phi _{0})\right)-1\right] +M_{L}\left[\cos \left( \frac{\beta
}{2}(\Phi _{B}-\phi _{L})\right)-1\right] \right) dt\ .
\end{eqnarray*}
Here, for finite \( M \)'s, 
Neumann boundary conditions are imposed in the underlying $c=1$ theory
 on the boundaries,
while if any of the \( M \) -s is infinite then the corresponding term
is absent and 
the boundary condition in the underlying conformal theory 
on that boundary is Dirichlet.
We can rewrite the Hamiltonian of the system in terms of the variables associated
to the plane using the map \((x,it)=\xi \to z=e^{i\frac{\pi }{L}\xi } \), and by
changing the integration variable we have
\be\label{n*}\eqalign{
H = & H_{CFT}+\frac{\mu }{2}\left( \frac{\pi }{L}\right) ^{2h_{\beta }-1}\int _{0}^{\pi }\left( V_{\beta }(e^{i\theta },e^{-i\theta })+V_{-\beta }(e^{i\theta },e^{-i\theta })\right) d\theta +\cr
 &\frac{M_{0}}{2}\left( \frac{\pi }{L}\right) ^{h_{\beta }}\left( e^{-i\frac{\beta }{2}\phi _{0}}\Psi _{\frac{\beta }{2}}(1)+e^{i\frac{\beta }{2}\phi _{0}}\Psi _{-\frac{\beta }{2}}(1)\right) +\cr
 &\frac{M_{L}}{2}\left( \frac{\pi }{L}\right) ^{h_{\beta }}\left(
e^{-i\frac{\beta }{2}\phi _{L}}\Psi _{\frac{\beta
}{2}}(-1)+e^{i\frac{\beta }{2}\phi _{L}}\Psi _{-\frac{\beta
}{2}}(-1)\right)\ .} 
\ee
 Now the computation of the matrix elements of the bulk and boundary
vertex operators $V_{\pm\beta}$ and $\Psi_{\pm\beta /2}$ (with
conformal dimension $h_\beta$) 
between the vectors of the appropriate conform 
Hilbert spaces is straightforward. Also the integrals can be calculated explicitly.
Truncating the Hilbert space at a certain conformal energy level
$E_{\rm cut}$ (which is nothing but the eigenvalue of the zeroth
Virasoro generator) and diagonalizing the
Hamiltonian numerically we arrive at the TCSA method. 

The TCSA Hamiltonian for BSG with Neumann boundary conditions at both
ends is obtained from (\ref{n*}) by setting $M_0=M_L=0$ and using
(\ref{X}), (\ref{X1}) with $n=1$, $r=\sqrt{4\pi}/\beta$ for $H_{CFT}$
and $V_\beta$, while the TCSA
Hamiltonian for BSG with mixed boundary conditions is obtained by
setting $M_0=M_L=0$ and using (\ref{3X}) and (\ref{3X1}) with the same
$n$ and $r$ for $H_{CFT}$ and $V_\beta$.
  
We choose our units in terms of the soliton mass $M$. The bulk
coupling $\mu $ is related to $M$ by
\be
\mu =\kappa (\beta)M^{2-2h_\beta},\qquad\qquad h_\beta
=\frac{\beta^2}{8\pi},
\ee
where $\kappa (\beta)$ is a dimensionless constant. In the bulk SG, from TBA
considerations, the exact form of $\kappa (\beta)$ was obtained in 
\cite{Zamm}, and we use the same form also here in BSG. Once we expressed
$\mu $, the Hamiltonian, (\ref{n*}), can be made dimensionless
$h=H/M$, depending only on  
the dimensionless volume parameter $l=ML$ and $\beta$ (in the general
case it also depends on $M_{0,L}/M$ and $\phi_{0,L}$).

\section{TCSA analysis of SGN in finite volume}

In this section we review the TCSA analysis of the BSG with Neumann
boundary condition. The aim of this investigation is twofold: first we
want to verify the various reflection factors and the boundary spectrum 
obtained from the bootstrap procedure, and second we want to obtain further 
information about the finite volume behaviour of SGN. We consider in
some detail the questions of the ground state and the associated
boundary energies, the low lying one particle states and their
reflections on the ground state wall and finally the new states
predicted by the bootstrap.

In SGN, as ${\bf C}$ symmetry persists, there are two sectors, the  
${\bf C}=1$ even and the ${\bf C}=-1$ odd ones. The bulk breathers naturally 
belong to one of them, as the ${\bf C}$ parity of the $n$-th breather is
$(-1)^n$. However, since solitons and anti solitons can reflect into 
themselves as well as into their charge conjugate partners, solitonic one 
particle states (i.e. states, whose energy and momentum are related by
$E=\sqrt{P^2+M^2}$ where $M$ is the soliton mass) are there in both sectors. 
The ground state is in the even sector, while we expect $\vert
1\rangle $ to be of the
lowest energy one in the odd sector (at least for $p<1/2$, which is true 
for all the cases investigated by us).

If at both ends of the strip Neumann boundary condition is imposed (NN case), 
then sometimes we have to face the problem of 
identifying the TCSA lines (states) corresponding to the 
symmetric/antisymmetric combinations of some identical excited states of the
identical boundaries, as described in section 4. 
Therefore in the NN case the finite volume spectrum 
and consequently the TCSA line sequences are more complex than the \lq naive' 
one derived by the bootstrap procedure in section 3. To avoid this 
complication when \lq verifying' the Neumann spectrum we consider a system, 
where Dirichlet boundary condition with $\phi_0=0$ is imposed on one of 
the boundaries, while keeping the Neumann one at the other (DN case). Since 
this mixed boundary condition is also ${\bf C}$ symmetric we do not lose the
presence of even and odd sectors on the one hand, while, on the other, we 
can be sure, that any boundary bound state (BBS) found in this case can be 
attributed to the Neumann end of the strip, as this Dirichlet boundary
condition has no 
bound states \cite{patr}. 

As charge conjugation acts on the fundamental scalar field by $\Phi\mapsto 
-\Phi$, it is straightforward to implement the projection onto the even and 
odd sectors in the conformal Hilbert spaces used in TCSA. This projection
has two beneficial effects: on the one hand it effectively halves the number 
of states below $E_{\rm cut}$, thus it drastically reduces the time needed 
to obtain the complete TCSA spectrum, and on the other the separate spectra 
of the even and odd sectors are less complex and therefore easier to study 
than the combined one.  

In our numerical studies $E_{\rm cut}$ varied between 16 and 25, and
this 
resulted in ${\rm 4.5}\times{\rm 10^3}$ - ${\rm 8}\times{\rm 10^3}$
conformal states per {\sl sectors}. In the DN case the number of
states below $E_{\rm cut}$ is independent of $p$, while in the NN case
it depends sensitively on it. 

\subsection{Ground state and boundary energy}

In any boundary field theory in finite volume, for large enough $L$-s, we 
expect on general physical grounds, that the volume dependence of the ground 
state energy can be written as
\be
E_0(L)=\epsilon_0L+{\cal E}_L+{\cal E}_R,
\ee
where $\epsilon_0$ is the ground state energy density and ${\cal
E}_{L,R}$ are the ground state  
boundary energies associated to the left and right ends of the strip. In the
BSG model one can argue that $\epsilon_0$ is nothing but the well known
bulk energy constant, $-\frac{M^2}{4}\tan(\frac{p\pi}{2})$, of the bulk SG 
theory. In BSG with Dirichlet boundary conditions, LeClair et al. \cite{Sal} 
were able to obtain ${\cal E}_{\rm Dir}$ from TBA, at least for those values of $p$ 
when not only the boundary but also the bulk scatterings become diagonal. The 
case of Neumann boundary condition is more complicated as there is no $p$ 
where the boundary scatterings would become diagonal.

\begin{figure}
\centering
\includegraphics{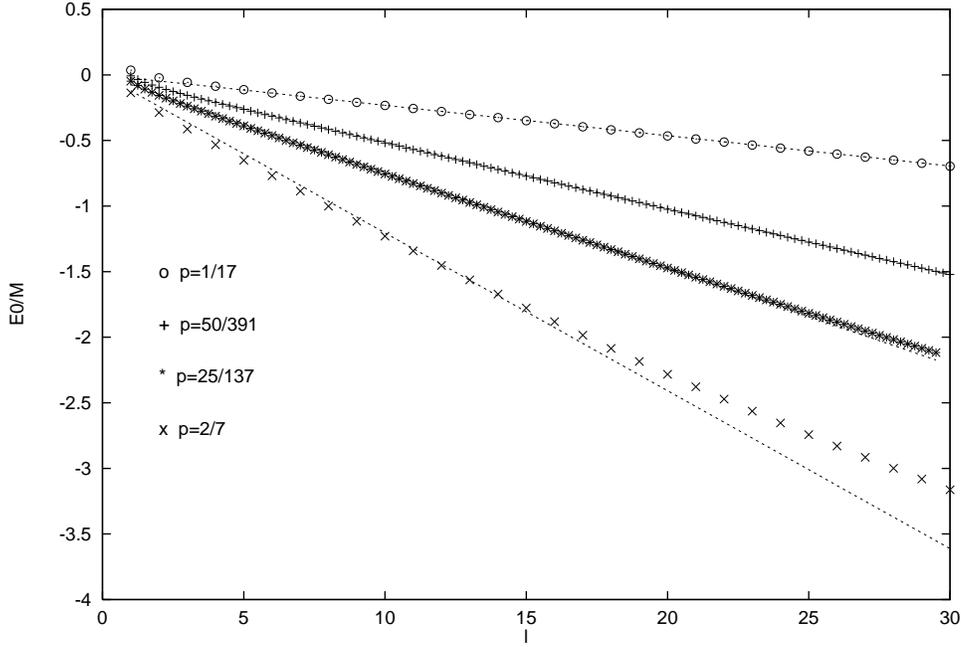}
\caption{Ground state energy versus $l$ in the DN case at various
$p$-s. The slope of the dashed lines is $-\frac{1}{4}\tan(\frac{p\pi}{2})$. }
\label{fig:dngs}
\end{figure}

\begin{figure}
\centering
\includegraphics{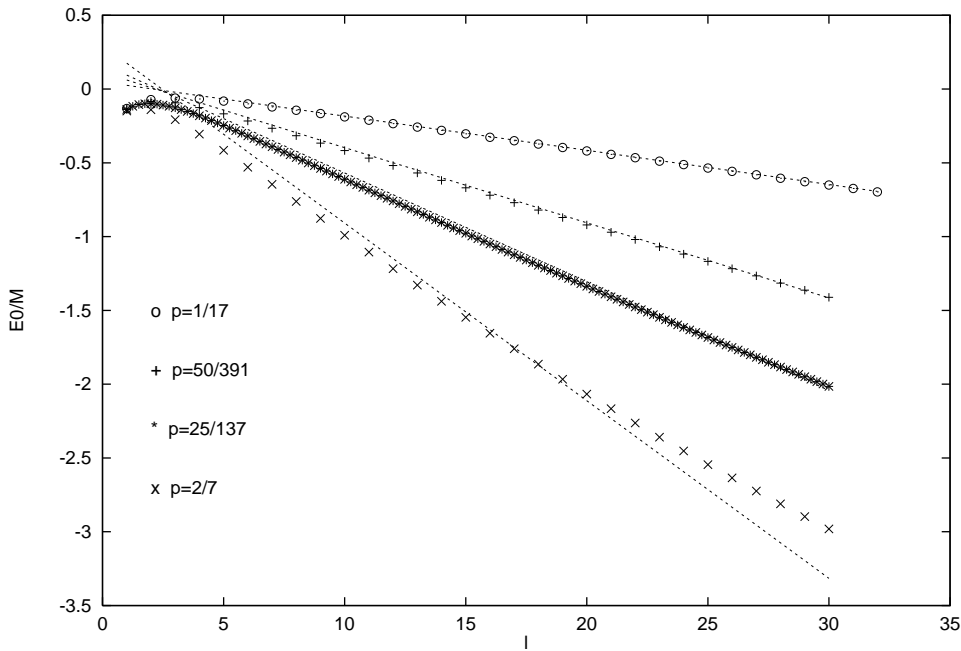}
\caption{Ground state energy versus $l$ in the NN case at various
$p$-s. The slope of the dashed lines is the same as on
Fig.(\ref{fig:dngs}), but their intersection is given by Eq.(\ref{iX}).}
\label{fig:nngs}
\end{figure}

Nevertheless using TCSA, we can get information on ${\cal E}_{\rm Neu}$. To this 
end we determined $E_0(l)/M$ ($M$ is the soliton mass and $l=ML$ is the 
dimensionless volume parameter) from our TCSA data at various values of $p$ 
both in the (DN) and in the (NN) cases; the results are shown on
Figs.(\ref{fig:dngs}-\ref{fig:nngs}) 
These
figures show in a convincing way that in these BSG models $\epsilon_0$ is 
indeed the expected one. Interestingly the (DN) data for $E_0(l)/M$ are 
consistent with a straight line passing through the origin; i.e. with
${\cal E}_{\rm Neu}(p)=-{\cal E}_{\rm Dir}(p,\phi_0=0)$. Since from ref.\cite{Sal} we know 
${\cal E}_{\rm Dir}$ analytically, using the (DN) data we have a prediction for 
$E_0(l)/M$ in the (NN) case: a straight line with slope parameter 
$-\frac{1}{4}\tan(\frac{p\pi}{2})$ and intersection 
\be\label{iX}
2\frac{{\cal E}_{\rm Neu}}{M}=-\frac{1}{2}\left( 1-\frac{2}{\cos(\frac{p\pi}{2})}+
\cot(\frac{\pi (p+1)}{4})\right)\ .
\ee
Fig.(\ref{fig:nngs}) shows that this prediction is consistent with our (NN) data.    

\subsection{Low lying one particle lines, reflections on the ground
state wall}

Next we summarize what we can learn about the Neumann spectrum by
studying the lowest lying one particle states.

\subsubsection{Breather lines}

Using the formalism developed in section 4 it is straightforward to
express in parametric form the Bethe - Yang (BY) lines corresponding to the various
breathers moving between ground state walls at the ends of the
strip\footnote{Among the breather lines these are of the lowest in energy
above the ground state.} :
\be\label{lpara}
(E(L),L)=(M_n\cosh\theta ,(2\pi I_n+i\log R^{(n)}_L(\theta )
+i\log R^{(n)}_R(\theta ))/(2M_n\sinh\theta ))\ .
\ee
Here $E(L)$ is the energy of the $n$-th breather above the ground
state, $\theta$ (the parameter of the line) is the rapidity of
$B^n$ (which, therefore, necessarily must be real and non
negative), and $M_n$ is the mass of $B^n$. $I_n$ is the quantum
number characterizing the line; if $R^{(n)}_L(0)R^{(n)}_R(0)=1$ then
$I_n$ is integer, while if $R^{(n)}_L(0)R^{(n)}_R(0)=-1$ then $I_n$ is
half integer. (If $R^{(n)}_L(\theta )=R^{(n)}_R(\theta )$ as in the
(NN) case, then only the first possibility appears, the second happens
e.g. in the (DN) case for the ${\bf C}$ even breathers).

\begin{figure}
{\centering \begin{tabular}{cc}
{\includegraphics{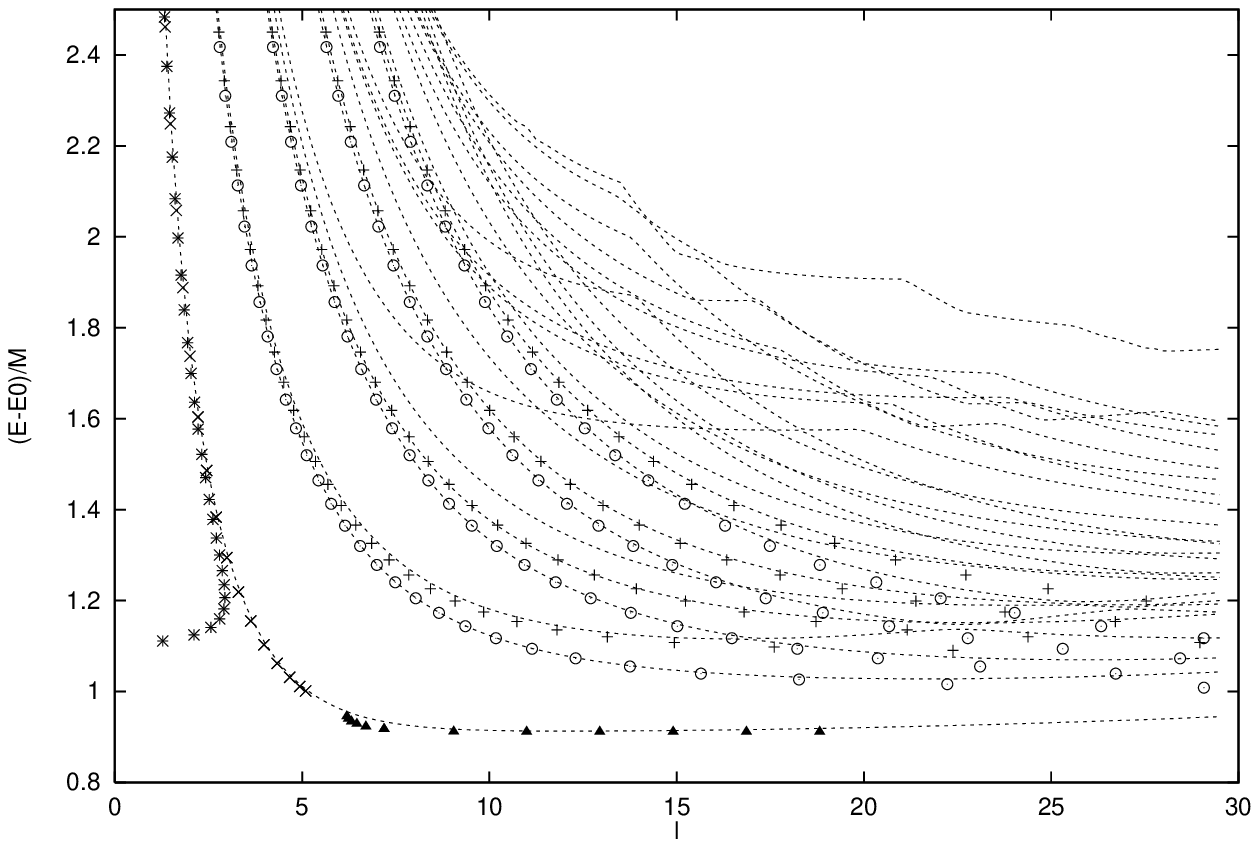}} \\
{\small DN spectrum: $+$ and $*$ denote the $B^2$ lines
with $I_{2}=\frac{1}{2}\dots \frac{7}{2}$ and
$-\frac{1}{2}$,}\\{\small 
$\circ$ and x the soliton lines
with $N=1\dots 4$ and $0$,}\\{\small the full triangles the $B^2$
line with imaginary rapidity. }\\
{\includegraphics{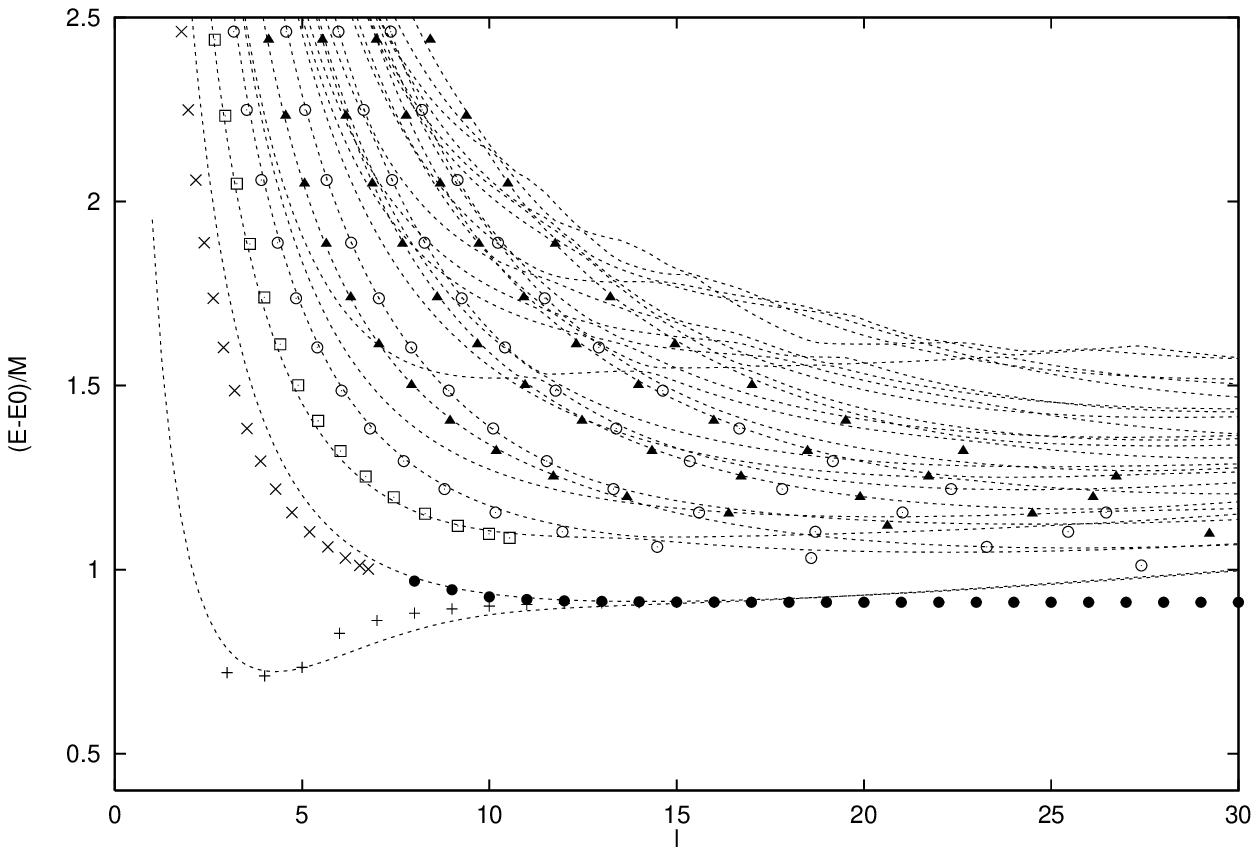}} \\
{\small  NN spectrum: the full triangles and empty boxes are the
$B^2$ lines with $I_2=1\dots 4$ and $0$,}
\\{\small 
$\circ$ and x the soliton lines
with $N=1\dots 4$ and $0$,}\\{\small the $\bullet$ and the $+$ the $B^2$
lines with imaginary rapidities just below and above $u_2$. }
\\
\end{tabular}\small \par}

\caption{TCSA data and soliton/breather Bethe Yang lines in the ${\bf
C}$ even sectors with \( p=25/137\).}\label{fig:symsec}
\end{figure}

The perfect agreement between the TCSA data and the predictions coming
from (\ref{lpara}) for $B^1$, $B^2$ and $B^3$ with $I_n>0$
in both the (DN) and the (NN) cases - as summarized in 
Figs.(\ref{fig:symsec}-\ref{fig:asymsec}) - gives
convincing evidence for the correctness of the reflection factors
given in \cite{gosh}. \footnote{On Figs.(\ref{fig:symsec}-\ref{fig:excited}) the
dimensionless energy levels above the ground state are plotted against
$l$. On all plots the continuous lines are the interpolated TCSA data
and the various symbols mark the data corresponding to the various BY
lines. Some of the higher TCSA lines appear to have been broken, the
apparent turning points are in fact level crossings with the other
line not shown. This happens because our numerical routine, instead of
giving the eigenvalues of the Hamiltonian in increasing order at each
value of $l$, fixes their order at a particular small $l$ and follows them
 -- keeping their order -- according to some criteria as $l$ is changing to higher values.} 

Using the parametric form of $L(\theta )$ given in (\ref{lpara})
together with the limits 
\[\lim_{\theta\to 0}R^{(n)}_L(\theta
)R^{(n)}_R(\theta )=\pm 1,\qquad \lim_{\theta\to\infty}\vert\log 
R^{(n)}_L(\theta )R^{(n)}_R(\theta )\vert <\infty\] it is
straightforward to show that for $I_n>0$ the range of $L(\theta )$ runs
from zero to infinity, as $\theta$ decreases from infinity to zero
(i.e. as we move from the UV to the IR). However, as it was explained in
section 4, in the case of $I_n=0$ there is a maximal value of $L_0$ beyond
which we can not go keeping $\theta$ real. We can see this
phenomenon on the $I_n=0$ lines on Figs.(\ref{fig:symsec}-\ref{fig:asymsec}). 
Interestingly, there are TCSA
lines on Figs.(\ref{fig:symsec}-\ref{fig:asymsec})
which, below a maximal $L$, can be described well by breather
(BY) lines with {\sl negative} quantum numbers like the $B^2$ line with
$I_2=-1/2$ or the $B^3$ line with $I_3=-1$. The essential
difference between the $L(\theta )$ functions with $I_n=0$ and the
ones with $I_n<0$ is that while the former ones reach their maximal value at
$\theta =0$, the latter ones do this at some positive
$\theta$. Therefore, while in the former case the BY lines can be continued
by going to purely imaginary rapidity, we can not do
this in the latter cases.   
As illustrated on Figs.(\ref{fig:symsec}-\ref{fig:all})
this continuation describes the TCSA data very well as $u=-i\theta$
moves from zero towards the first singularity in the reflection
factors, if this first singularity corresponds to a bound state pole
in bootstrap: see the $B^1$ lines ending in the $e_1$ bound state
in both the (DN) and the (NN) cases. 

These Figures also show that the idea put forward in section 4 to
describe (at least for large enough $L$-s) the TCSA lines corresponding
to BBS by appropriate one particle (here breather) BY lines with
purely imaginary rapidities (independently whether they are obtained
from continuation through $\theta=0$ or not) works indeed. At this
point the comparison of the DN and NN spectra is very instructive:
while in the DN cases there are only single lines that descend to
$e_1$ or $e_2$ from above for large $L$-s, in the NN cases one can see both an
increasing and a decreasing line tending to $e_1$ and $e_2$. In the
latter case these nearly degenerate TCSA states may be interpreted as
the symmetric and antisymmetric wave functions in which one of the identical
(and hence indistinguishable) boundaries is in 
an excited state and the other one is in the ground state. For large
enough $L$-s the volume dependence of the nearly degenerate energies
is well described by BY lines with purely imaginary rapidities just
above or just below the values corresponding to $e_1$ or $e_2$.  

On the last two plots in Fig.(\ref{fig:all}) we also exhibit the
appearance of the $\vert 3\rangle$ state in an expected way. 
For $p=25/137$ (when the state $\vert 3\rangle$ is absent) the BY line
marked by the $\circ$-s is a solitonic line 
with $N=1$ extending to infinity, while for $p=50/391$ (when $\vert
3\rangle$ is already present) it is a solitonic line with $N=0$,
ending at a maximal $l$. In this latter case the TCSA data in the
continuation of this BY line are correctly described by the x-s,
which are the data from the BY line of
the third breather with imaginary rapidity just below $u_3$.

\subsubsection{Soliton lines}

As mentioned in the introduction to this section in the description of
solitonic one particle states we have to face the problem of non
diagonal reflections. In the ${\bf C}$ symmetric case (i.e. when
$s\leftrightarrow \bar{s}$ is a symmetry) the one particle soliton
anti soliton transfer matrix is $2\times 2$, and the Bethe Yang
equations using the eigenvalues of this matrix take the form
\be
1=e^{i2ML\sinh\theta}(P_L(\theta )\pm Q_L(\theta ))(P_R(\theta )\pm
Q_R(\theta ))\ ,
\ee
where, in the ${\bf C}$ even/odd sectors the upper/lower signs apply,
and $P_{L,R}(\theta )$ ($Q_{L,R}(\theta )$) denote the soliton soliton
(soliton anti soliton) reflection factors on the ground state left and
right boundaries. In the (NN) case these
equations simplify to
\be\label{nnsol}
1=e^{i2l\sinh\theta}(P(\theta )\pm Q(\theta ))^2,\qquad l=ML,
\ee
while in the (DN) case - when on the Dirichlet wall $Q$ vanishes - to
\be\label{dnsol}
1=e^{i2l\sinh\theta}(P(\theta )\pm Q(\theta ))P_D(\theta ).
\ee
(Here $P_D$ is the $\xi=0$ case of the soliton reflection factor in 
BSG with Dirichlet boundary conditions: $P_D(\theta )=P^{(+)}(\theta
)\vert_{\xi=0}$ of ref. \cite{patr}).  Using the infinite product form
of $P$, $Q$ and $P_D$ it is straightforward to verify, that the
expressions multiplying $e^{i2l\sinh\theta}$ in
Eq. (\ref{nnsol}-\ref{dnsol}) are indeed pure phases -- at least for
$0\le \theta\le\infty$ -- thus these equations make sense. 

\begin{figure}
{\centering \begin{tabular}{cc}
{\includegraphics{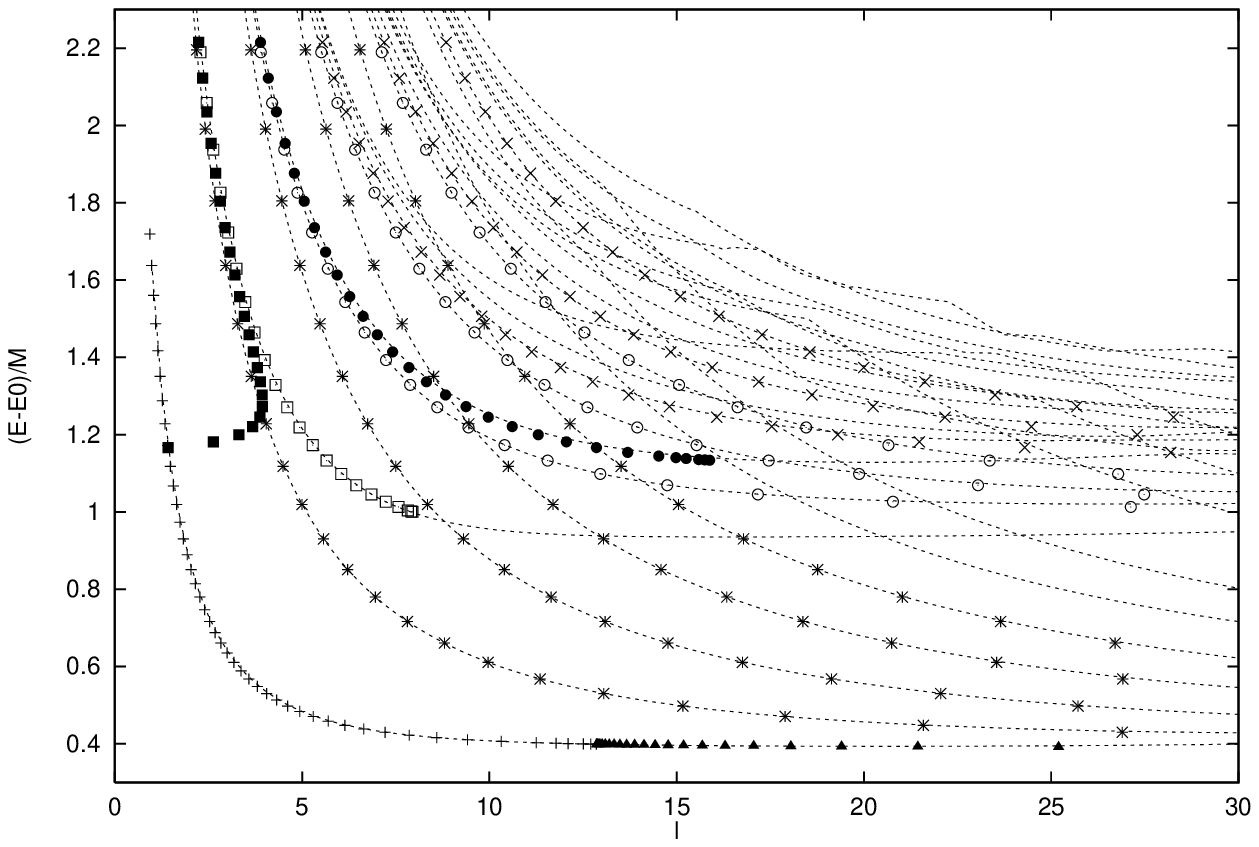}} \\
{\small DN spectrum: $*$ and $+$ denote the $B^1$ lines
with $I_{1}=1\dots 4$ and
$0$,
$\circ$ and the empty boxes}\\{\small the soliton lines
with $N=1\dots 3$ and $0$, the full triangles the $B^1$
line with imaginary rapidity,}
\\{\small the x, $\bullet$, and the full boxes the $B^3$
lines with $I_3=1\dots 3$, $0$, and $-1$.}
 \\
{\includegraphics{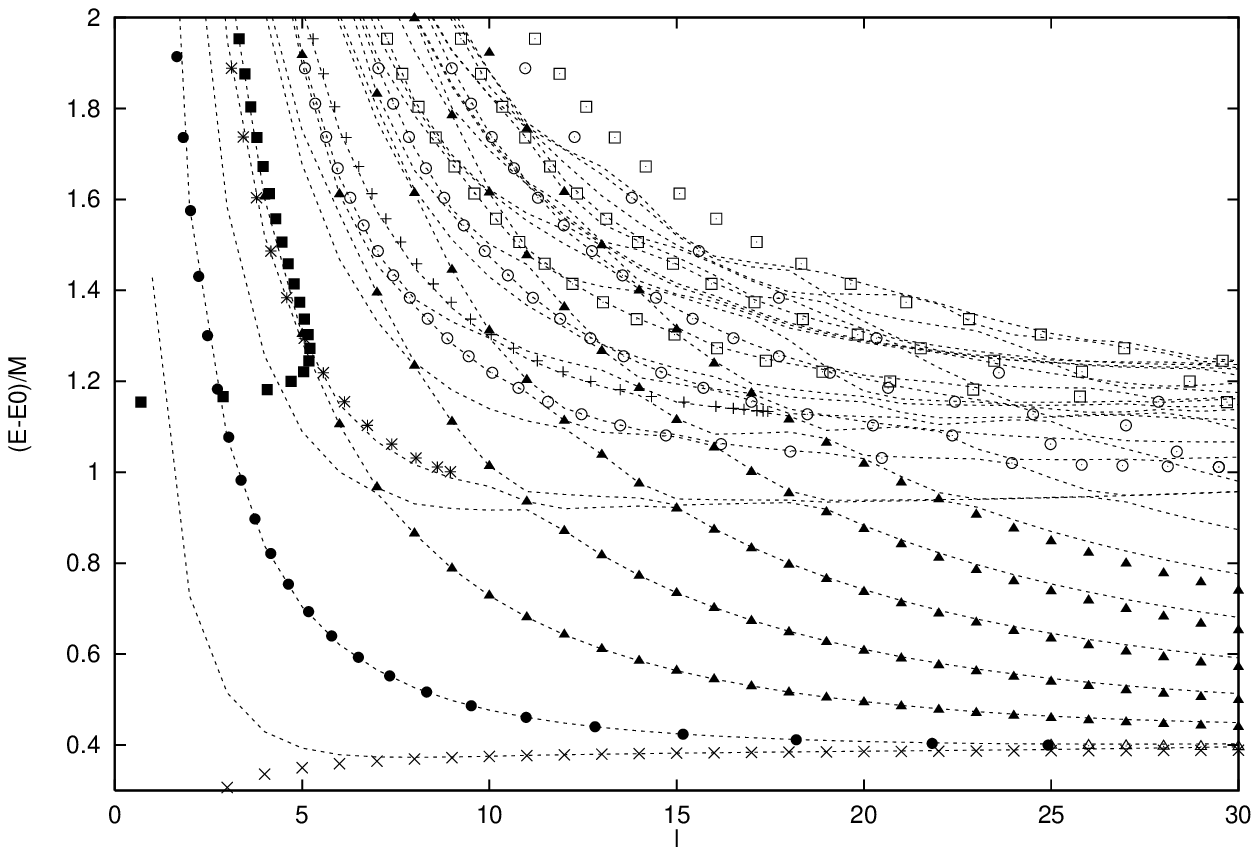}} \\
{\small  NN spectrum:  full triangles and $\bullet$ denote the $B^1$ lines
with $I_{1}=1\dots 5$ and
$0$,
$\circ$ and $*$}\\{\small the soliton lines
with $N=1\dots 4$ and $0$, x and $\Delta$ the $B^1$
lines with imaginary rapidities just}\\{\small above and below $u_1$,
the empty and full boxes and $+$  the $B^3$
lines with $I_3=1\dots 3$, $-1$, and $0$.
}\\
\end{tabular}\small \par}

\caption{TCSA data and soliton/breather Bethe Yang lines in the ${\bf
C}$ odd sectors with \( p=50/391\). }\label{fig:asymsec}
\end{figure}

The infinite product representation is not very useful when we compare
BY lines coming from Eq. (\ref{nnsol}-\ref{dnsol}) to the TCSA
data. Therefore, using the well known integral representation for the
logarithm of the Gamma function, we recast $l(\theta )$ such that the
parametric form of the solitonic BY lines looks like (with $E(l)/M$
denoting the energy above the ground state):
\be\label{int1}
(E(l)/M,l)=(\cosh\theta ,l(\theta )),
\ee
where
\be\label{int2}
l(\theta )=\frac{N\pi+{\bf a}i\log r_\pm +I(p,\theta)}{\sinh\theta},
\ee
with
\be\label{int3}
I(p,\theta )=\int\limits_{0}^{\infty}\frac{dy}{y}\sin(\frac{2\theta
y}{\pi})\Bigl(
\frac{2\sinh(3y/2)\sinh([1-p]y/2)}{\sinh(py/2)\sinh(2y)}+
\frac{{\bf a}\sinh(py)-\sinh(y)}{\cosh(y)\sinh(py)}\Bigr)\ ,
\ee
and
\be\label{int4}
r_+=\frac{\sin\Bigl(\frac{\pi}{4p}-i\frac{\theta}{2p}\Bigr)}{\sin\Bigl(\frac{\pi}{4p}
+i\frac{\theta}{2p}\Bigr)},\qquad\qquad  
r_-=\frac{\cos\Bigl(\frac{\pi}{4p}-i\frac{\theta}{2p}\Bigr)}{\cos\Bigl(\frac{\pi}{4p}
+i\frac{\theta}{2p}\Bigr)},
\ee
in the ${\bf C}$ even/${\bf C}$ odd sectors. 
Here 
the parameter ${\bf a}$ is one,  ${\bf a}=1$, in the (NN) case and
is one half, ${\bf a}=1/2$, in the (DN) case; the quantum number $N$, 
characterizing the BY line, is a non negative integer in all sectors
and all cases and -as before - $\theta$ is both the parameter of the
line and the rapidity of the solitonic particle. (Although this representation
of the solitonic BY lines works for real rapidities only, one can show
that all the \lq\lq boundary dependent'' poles at $\theta=iu$ come
from $\log r_\pm$, in fact from the set of poles, 
$\theta=i(\frac{\pi}{2}-kp\pi)$,  
$r_+$ has poles for $k=2N$ while $r_-$ for $k=2N+1$). 

On Figs.(\ref{fig:symsec}-\ref{fig:asymsec})
we demonstrate the excellent agreement between the TCSA data and
the BY lines given by (\ref{int1}-\ref{int4}) with $N>0$ in both the
${\bf C}$ even and the ${\bf C}$ odd sectors of the (DN) and (NN)
cases. This agreement gives a strong evidence for the correctness of
the solitonic reflection factors given in \cite{GZ}.
 
Note that in some cases we have TCSA lines that below a maximal $L$
can be described well by solitonic BY lines with zero quantum number,
$N=0$. The continuation of these BY lines to imaginary rapidities --
that worked for the breather lines -- is hampered by the fact, that in
all the cases investigated, the pole in $(P\pm Q)^2$ or $(P\pm Q)P_D$
nearest to the origin turned out to be a Coleman-Thun pole. Therefore
the solitonic BY lines continued to imaginary rapidities deviate from
the TCSA data.

\subsection{New boundary bound states and reflections on excited
walls}

\begin{figure}
{\centering \begin{tabular}{cc}
\resizebox*{!}{0.4\textwidth}{\includegraphics{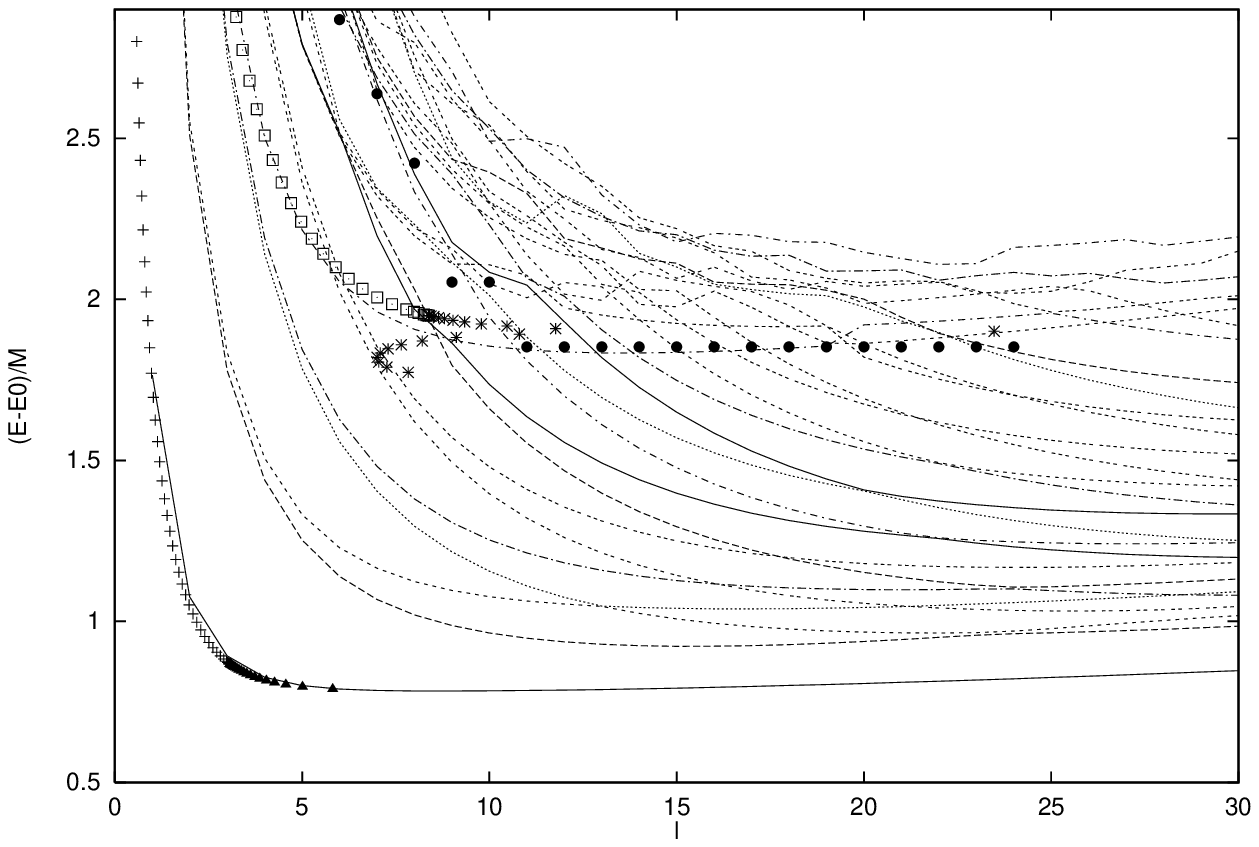}} \\
{\small \( p=2/7 \), no \(\vert 1,2\rangle\) state exists}\\
\resizebox*{!}{0.4\textwidth}{\includegraphics{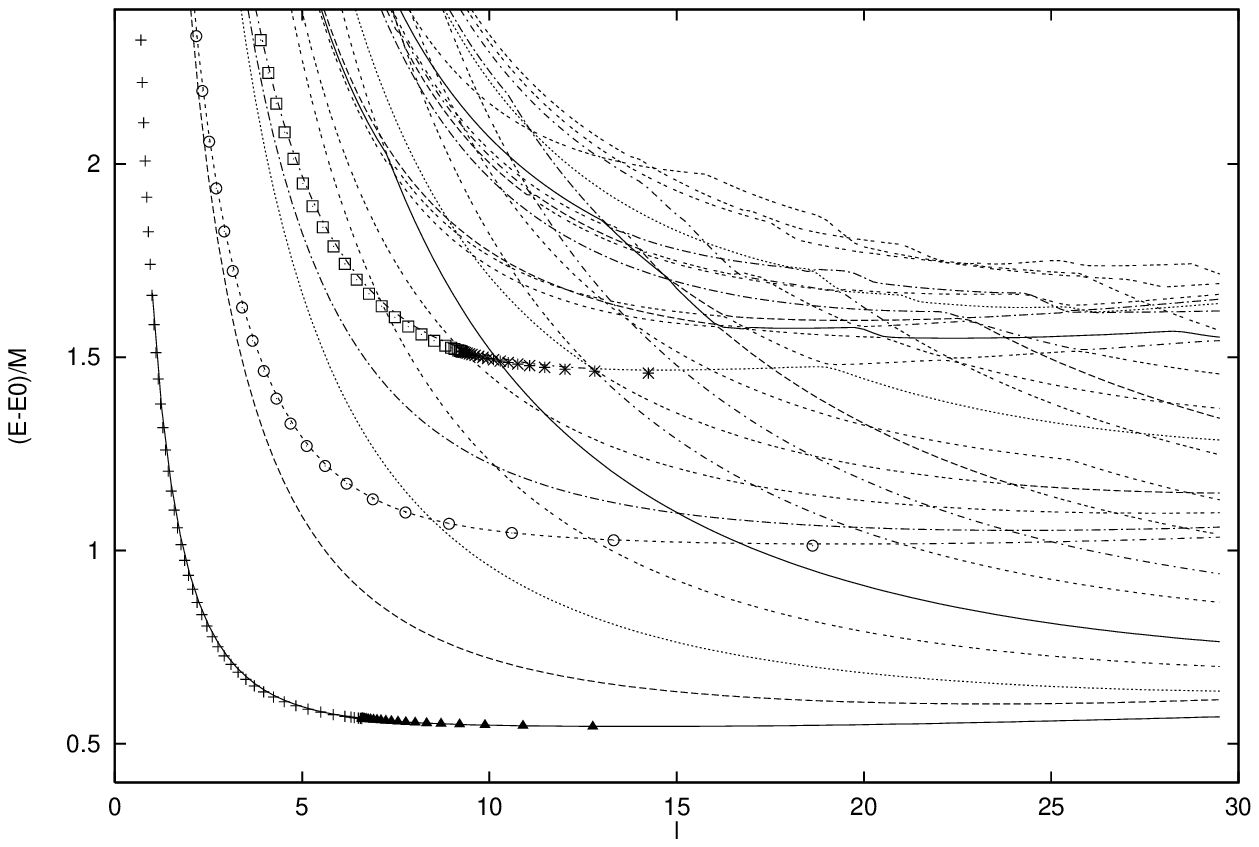}} \\
{\small \( p=25/137 \), the \(\vert 1,2\rangle\) state exists}\\
\resizebox*{!}{0.4\textwidth}{\includegraphics{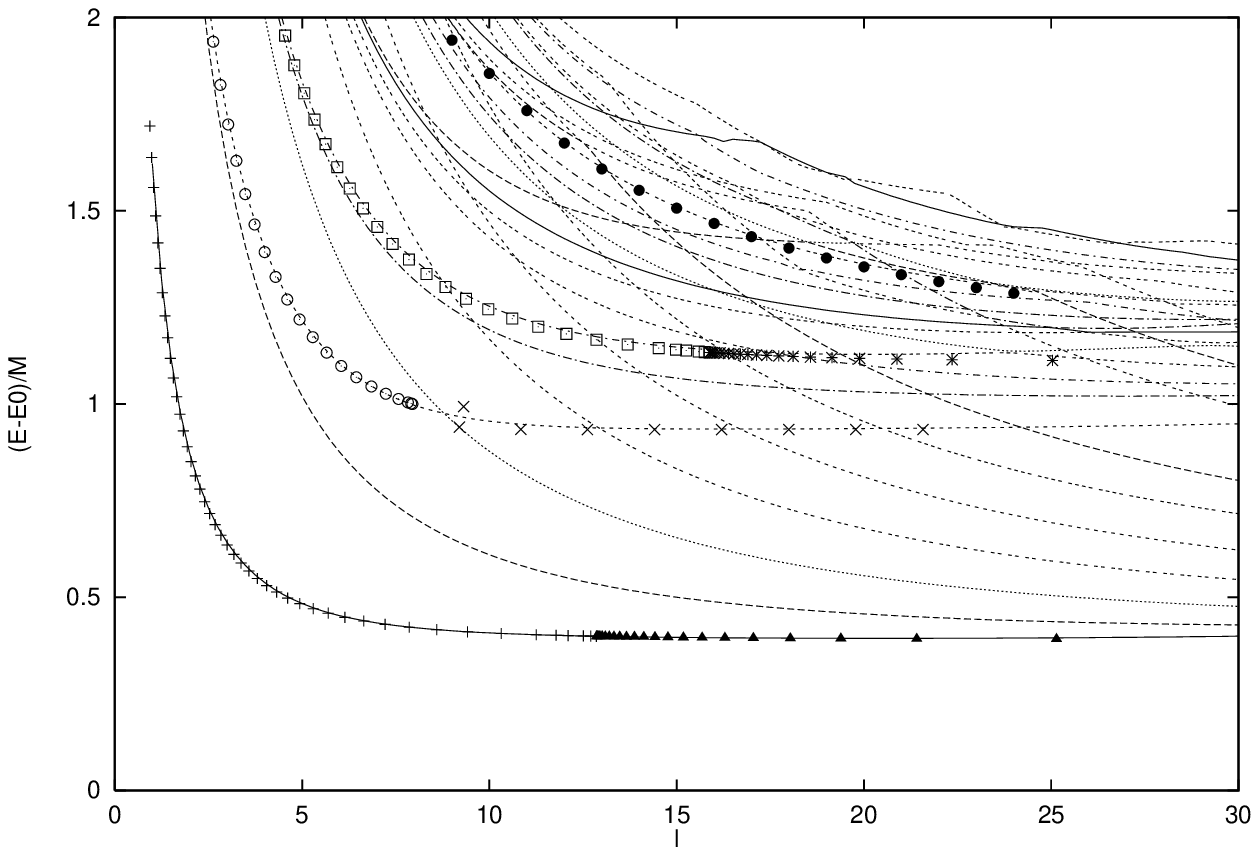}} \\
{\small \( p=50/391 \), the \(\vert 1,2\rangle\) state exists}\\
\end{tabular}\small \par}

\caption{: Evidence for the existence of the $\vert
1,2\rangle $ state from three ${\bf C}=-1$ DN spectra.  }\label{fig:all}
\end{figure}

The most interesting application of TCSA is to find evidence for the
existence of the new BBS-s predicted by the bootstrap and check the
correctness of the \lq excited' reflection factors, i.e. the ones
describing the scattering of breathers/solitons on the wall in an
excited state. In this inquiry it is very helpful to analyze the DN
spectra: if, e.g. we find a TCSA line (state) with asymptotic (large
$l$) energy $e_1+e_2$ in the DN spectrum, then we can be sure that it
indeed corresponds to the $\vert 1,2\rangle $ state of the Neumann
boundary. (Note that in the NN case there are more states having this
asymptotics, as the energy of the configurations, when one of the
Neumann boundaries is in the $\vert
1\rangle $ state and the other is in the $\vert 2\rangle $ state, also
tends to $e_1+e_2$). 

The state $\vert 1,2\rangle $  can be generated by the third breather, furthermore, 
the corresponding  pole at $u_1$ is the one nearest to the origin in the
third breather's reflection factor. Therefore it is expected that the
state $\vert 1,2\rangle $ becomes visible as the imaginary
continuation of a $B^3$ BY line with $I_3=0$.
 We collected the evidence for the existence of $\vert 1,2\rangle $ on 
fig.(\ref{fig:all}).  On all three plots the continuous lines are the interpolated
TCSA data, the $+$-s and the empty boxes denote the BY line of $B^1$ and
$B^3$ with $I_{1,3}=0$ respectively, while the full triangles and the
$*$-s denote the continuations of these lines to purely imaginary
rapidities with $u$ ranging from zero to $u_1$. Note, that 
for $p=2/7$, when, according to bootstrap
considerations, there is no $\vert 1,2\rangle $ state and the
imaginary continuation of the $B^3$ BY line should run into a Coleman -
Thun pole, the BY line indeed departs from the TCSA data, (which, in this
case are well described by a $B^2B^1$ BY line where the second
breather's rapidity is imaginary and the first breather's quantum
number is $I_1=1$, the data corresponding to this line are given by
the $\bullet$-s). On the other hand, when $p$ is such that $\vert 1,2\rangle $
 should exist, we see that the imaginary continuation of the third 
breather's BY line describes the TCSA data just as well as that of the
first breather's line, which runs into $e_1$. 

\begin{figure}
\centering
\includegraphics{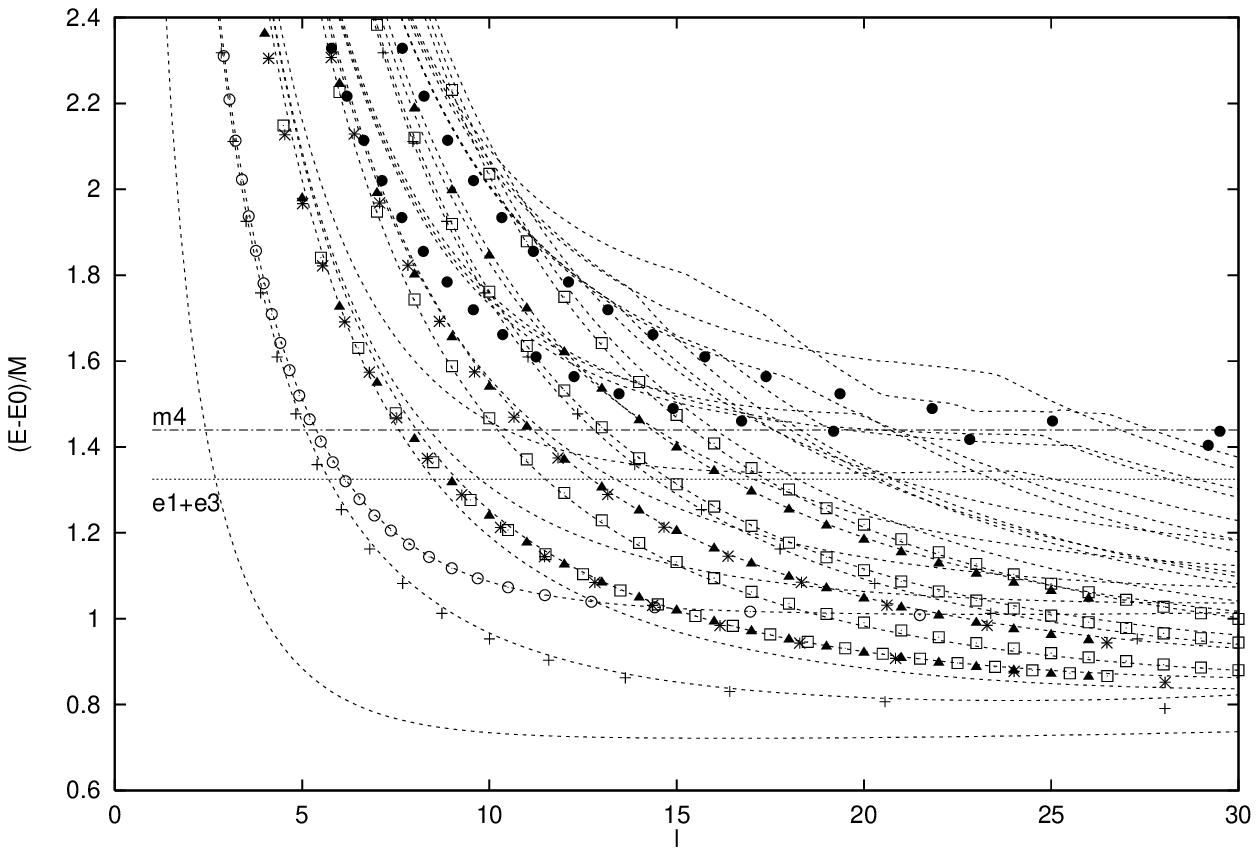}
\caption{TCSA data, the state $\vert 1,3\rangle $, and \lq\lq
excited'' BY lines in the ${\bf C}$ even sector of the DN model with $p=50/391$. }
\label{fig:excited}
\end{figure}

Further evidence for the existence of a new state predicted by the
bootstrap is given on Fig.(\ref{fig:excited}), where we exhibit a
single TCSA line descending to $e_1+e_3$. In the light of what we said
above this indicates the existence of the boundary bound state $\vert 1,3\rangle $.   
(The horizontal lines corresponding to $m_4=M_4/M$ and the
dimensionless $e_1+e_3$ are drawn
to guide the reader's eye to distinguish this line from the real fourth
breather's BY lines lying above $m_4$.) This state corresponds to the
pole at $u_2$ in the fourth breather's
reflection factor on the ground state boundary. Therefore, above a
certain minimal $l$, the TCSA data in this line can certainly be
described as the BY line of $B^{(4)}$ with imaginary rapidity below
$u_2$, but, 
since this reflection
factor also has a pole at $u_1$, can not be obtained as the analitycal continuation of a BY
line through $\theta =0$.

The bootstrap procedure gave not only the spectrum of the BBS but also
the reflection factors describing the scatterings of the various
breathers/solitons on excited boundaries. On Fig.(\ref{fig:excited})
we also confirm the correctness of these new reflection factors. 

There are two ways one can describe the scattering of a breather,
($B^n$ say), on an excited boundary $\vert k\rangle $. The first
is to use a BY line (\lq\lq excited'' BY line) following from
Eq.(\ref{scalar_quant_cond})
\be\label{elpara}
(E(L),L)=(e_k+M_n\cosh\theta ,(2\pi I_n+i\log R^{(n)}_L(\theta )
+i\log R^{(n)}_R(\theta ))/(2M_n\sinh\theta ))\ ,
\ee
where $R^{(n)}_R$ or $R^{(n)}_L$ now corresponds to the appropriate    
$R^{(n)}_{\vert k\rangle}$, and the energy above the ground state is given by 
$e_k+M_n\cosh\theta $. This description, however, ignores the fact,
that, in finite volume $L$, even in the absence of the moving
$B^n$, the energy of the state $\vert k\rangle $ only
asymptotically coincides with $e_k$. To remedy this we may try to use a two
particle BY line, where ground state reflection factors appear, but
one of the particles moves with imaginary rapidity, (which is in the
vicinity of a pole corresponding to $\vert k\rangle $ in the
appropriate reflection factor). It is expected that in this way we
describe more accurately the volume dependence of the energy of the
state 
combined from the moving
breather and $\vert k\rangle $.

Since the TCSA data are usually the better the lower the level is, we
may hope to confirm the new reflection factors mainly for light particles
reflecting on lowly excited boundaries. In the $p$ range we investigated
the first breather is the lightest particle, therefore on  
Fig.(\ref{fig:excited}) we compare $B^1$'s BY line on the first
excited boundary $\vert
1\rangle $, and the TCSA data. The points marked by $*$ correspond to
$I_1=1,\, 2$ in (\ref{elpara}). The data, corresponding to the alternative description, using  
two particle $B^1B^1$ lines with one imaginary rapidity and one real one
with quantum numbers $I_1=1\dots 3$ are marked by the full triangles;
they give a somewhat better description of the TCSA lines. The four
lines marked by the empty boxes correspond to two particle  $B^1B^1$ lines
with real rapidity and quantum numbers $(1,0)$, $(-1,1)$, $(1,2)$ and
$(-2,2)$ respectively. The lines marked by $\circ$ and $+$ are soliton
with $N=1$ and $B^2$ on ground state boundary with
$I_{1/2}=\frac{1}{2}$, and $I_{1/2}=\frac{7}{2}$, respectively; their
function is to help the reader to identify the already known breather/soliton      
 ground state BY lines. Finally, for comparison, we also included in
this plot two lines - marked by $\bullet$-s;  they are solitonic
\lq\lq excited'' BY lines with $N=1,2$. 

Summarizing, we can say, that
for the first breather, 
the satisfactory agreement we find between the TCSA data and the various ways of
describing its scattering on the first excited boundary
gives support for the correctness of the new reflection factors. We
expect that the
relatively poor agreement between the (considerably higher) TCSA data
and the \lq\lq excited'' solitonic BY lines could be improved by
replacing the latter by two particle soliton-$B^1$ lines where the
breather is moving with an imaginary rapidity.   
On
Fig.(\ref{fig:excited}) we also show examples of real two particle BY 
lines correctly describing TCSA data as well as instances, when a TCSA line is
described by several BY lines. In the ${\rm C}$ odd sector the lowest
lying `\lq\lq excited'' BY lines describe the scattering of $B^2$
on $\vert 1\rangle $; we repeated also in this case the analysis
described above and obtained the same qualitative conclusions.   

\section{Conclusions}

In this paper we analyzed the sine-Gordon model with Neumann boundary
condition (SGN). In particular we established the spectrum of boundary
states, developed a framework to describe finite size effects
in boundary theories and used this framework together with TCSA to
confirm the boundary  states and the reflection factors.

Of the boundary states we showed that they can be labelled by an
increasing sequence of positive integers $n_i$ satisfying 
$n_i<\left[\frac{\lambda}{2}\right]$, and denoted as $\vert n_1,\dots
n_k\rangle$. For the states $\vert n\rangle$ we showed that their
energies coincide with that of the WKB quantized classical boundary
breather. We realized the existence of the states $\vert n,k\rangle $
by studying on the one hand the poles of the breather's reflection
factors on non excited boundaries, and on the other, the poles of the
soliton reflection factor $P_{\vert n\rangle }(u)$ on the excited
state $\vert n\rangle $. The novel feature of this study was that we 
computed also the residues of the Coleman-Thun diagrams and the
reflection factors. This way we discovered that Coleman-Thun diagrams 
and creation of boundary bound states may coexist: this happens if the
residue of the diagram is not sufficient to explain the residue of the
reflection factor. Finally we proved that the new 
boundary states labelled by more than one integers can be generated 
successively by appropriate soliton
reflections. We also determined all solitonic and breather reflection 
factors on these new states and explained their poles.

It is interesting to compare the labeling of boundary states here in
SGN to that of in sine-Gordon model with Dirichlet boundary
conditions \cite{patr}. In the latter case the labeling is more
complicated: it consists of two sequences of positive integers. The
explanation is that in the Dirichlet case, in contrast to the
Neumann one, the solitons and the anti solitons generate different
boundary states.  

Moving from the infinite half line to a finite line segment, 
we gave the general form of Bethe-Yang equations for particles in
finite volume with boundaries and used them to discuss the appearance 
of boundary bound states in this finite size setting. We showed in
particular, that these states can be described as solutions of the BY
equations with purely imaginary rapidity.

Finally we studied the finite volume spectrum of SGN by the truncated conformal
space approach. We established, that the ground state energy density
of this model coincides with that of the bulk SG and could also relate
the ground state Neumann boundary energy to that of the Dirichlet one
with $\phi_0=0$. The perfect agreement we found between certain lines
in the TCSA spectrum and the one particle soliton/breather Bethe-Yang
lines gives evidence that the reflection factors entering into the BY
lines are indeed correct. We showed that the boundary bound states
indeed appear in the TCSA spectra as predicted by the Bethe-Yang
equations with purely imaginary rapidity. Finally in the TCSA data we
found evidence for the existence of the new multi labelled states and
also for the excited state reflections.

\begin{center}\textbf{Acknowledgements}\end{center}

We thank F. W\'agner for his assistance in the early stages of the
numerical work. G. T. thanks the Hungarian
Ministry of Education for a Magyary Postdoctoral Fellowship.  
This research was supported in part by the Hungarian
Ministry of Education under FKFP 0178/1999, 0043/2001 and by the Hungarian National
Science Fund (OTKA) T029802/99.

\appendix

\makeatletter
\renewcommand\theequation{\hbox{\normalsize\Alph{section}.\arabic{equation}}}
\makeatother                                                                  \makeatletter
\renewcommand\thefigure{\hbox{\normalsize\Alph{section}.\arabic{figure}}}
\makeatother                                                                    
\section{The proof of the existence of boundary bound states }

In this appendix we prove that the boundary bound state \( |n_{1},n_{2},\dots ,n_{l},k\rangle  \)
can be created by reflecting a soliton with rapidity \( \nu _{k} \)
on the boundary state \( |n_{1},n_{2},\dots ,n_{l}\rangle  \). In
doing so we show that no on-shell diagram exist for this process.
Since any on-shell diagram consists of virtual processes we omit the
attribute virtual from now on. One possible diagram is shown on 
 Fig.(\ref{fig:top line}).

\vspace{0.3cm}
{\centering
\begin{figure}~~~~~~~~~~~~~~~~~~~~\resizebox*{11cm}{6cm}{\includegraphics{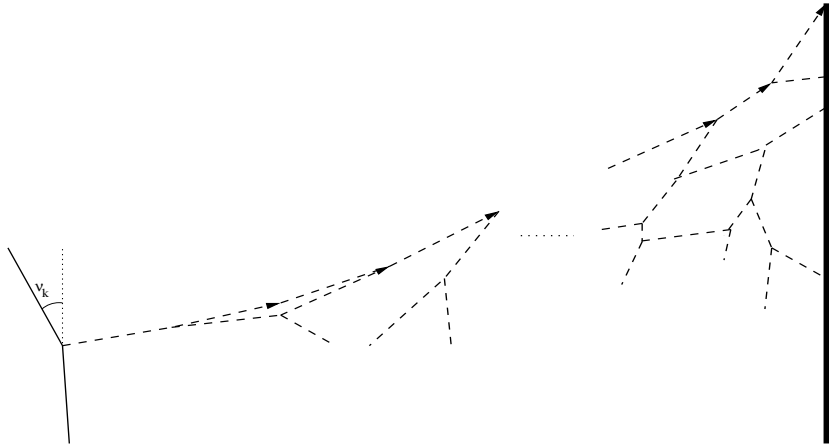}} 
\caption{: Schematic draw of the top line}\label{fig:top line}\end{figure}\par}
\vspace{0.3cm}

The leftmost process in which a soliton arrives with rapidity \( \nu _{k} \)
and virtually fuses with a breather into a soliton is called the final
process while the breather is the final breather. Now if we follow
the breather line towards the wall and turn in each 3-point vertex
to the left we obtain a line which we call the top line. (In case of
a 4-point vertex, that describes a possible two particle scattering
instead of fusion, we follow the line with the same rapidity). The
top line reaches the wall in the initial breather\footnote{%
We call it in this way, because in our convention, when time flows
from top to bottom, the emission of this breather by the wall precedes
all the other processes on the top line. 
}. Of course we have analogous diagrams by replacing some of the breather
lines with soliton lines. In this case the final/initial soliton terminology
is used. Clearly in the final process the particle created must travel
towards the wall i.e. it must have positive rapidity. 

\begin{figure}~~~~~~~~~~\subfigure[Final breather
]{\resizebox*{!}{6cm}{\includegraphics{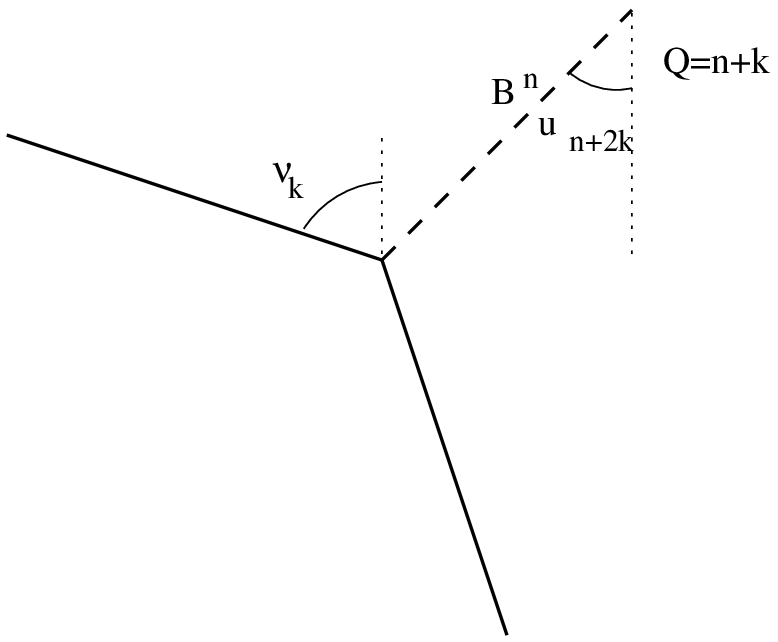}}}
~~~~~~~~~~~~~~~~~~~\subfigure[Final
soliton]{\resizebox*{!}{6cm}{\includegraphics{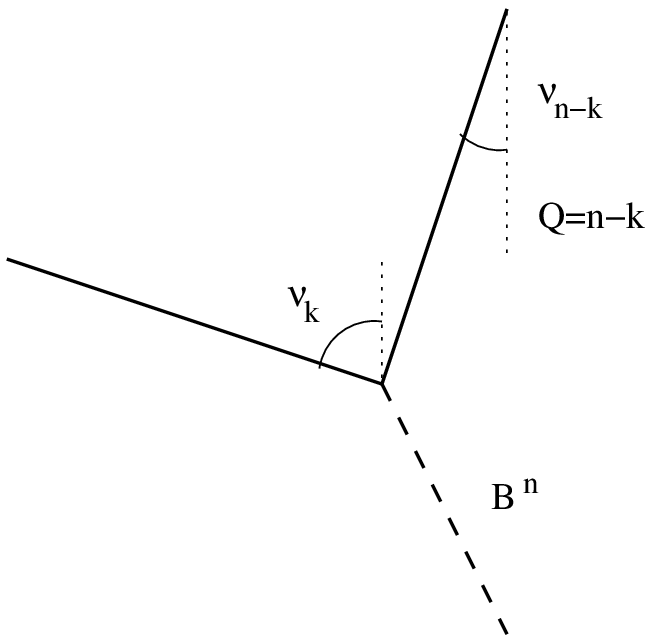}}}\caption{: Final
processes}\label{fig:final}\end{figure} 

The outline of the proof goes as follows: First of all we introduce
the notion of {}``proper{}'' particles and show that particles with
rapidity suitable for being final ones are {}``proper{}''. Since
any initial particles are also {}``proper{}'' we show that in the
diagram above only {}``proper{}'' particles exist. Moreover, we
determine all the bulk processes that take place on the top line and
call them {}``proper{}'' processes. Having introduced a charge like
quantity for {}``proper{}'' particles we show that the maximal charge
of the initial particles is \( n_{l} \). Since the charge never increases
in {}``proper{}'' processes and the charge of the final particle
is at least \( k \) which is larger than \( n_{l} \) we conclude
that no on-shell diagram exists. 

Let us go into the details. We call a breather {}``proper{}'' if
it has rapidity \( -u_{n}\, ,\, \, n\in N \) and a soliton is {}``proper{}''
if its rapidity is \( -\nu _{n} \). The possible final processes
are shown by diagram (a-b) on Fig.(\ref{fig:final}). Clearly both the final breather and the
final solitons are {}``proper{}''. Initial particles are produced
by the decay of boundary bound states, the various virtual processes
are described on the three diagrams of Fig.(\ref{fig:init}). Clearly all the initial particles are
also {}``proper{}''. For {}``proper{}'' particles we introduce
the notion of charge. For \( B^{n} \) with rapidity \( u=-u_{m} \)
the charged is defined to be \( Q=\frac{n+m}{2} \) while for a soliton
with rapidity \( u=-\nu _{k} \) it is simply \( Q=k \). The charges
of the initial particles are also displayed on Fig.(\ref{fig:init}). The maximum
of the initial charge is \( n_{l} \), obviously. The final charges are
 also displayed on Fig.(\ref{fig:final}). Since the particle,
produced in the final process, has to travel towards the wall, (i.e.
\( \nu _{n-k}<\nu _{k} \) for diagram (b)), we conclude that the
final charge must be larger then \( k \). Now we analyze how the
charges change on the top line. 

%\vspace{0.3cm}
{\centering \begin{figure}\subfigure[Breather initial
state]{\resizebox*{!}{4.8cm}{\includegraphics{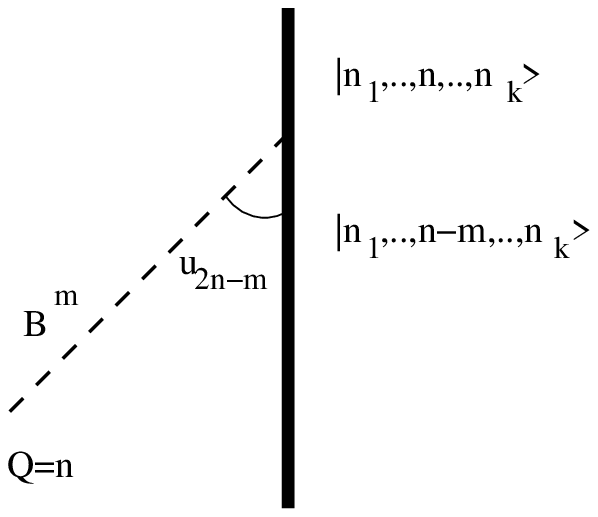}}}
\subfigure[Breather initial
state]{\resizebox*{!}{4.8cm}{\includegraphics{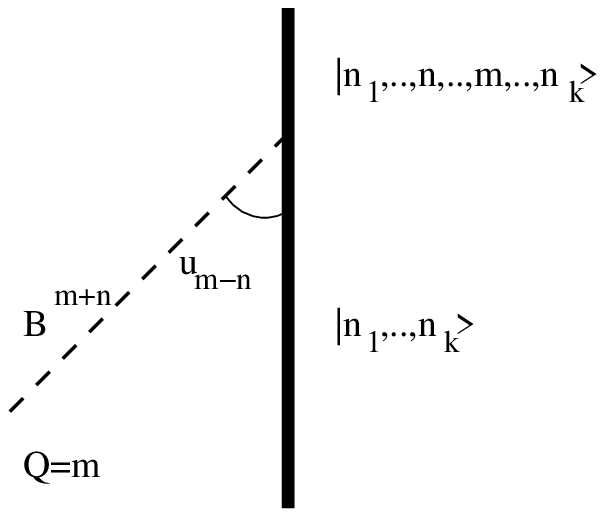}}}
\subfigure[Soliton initial
state]{\resizebox*{!}{4.8cm}{\includegraphics{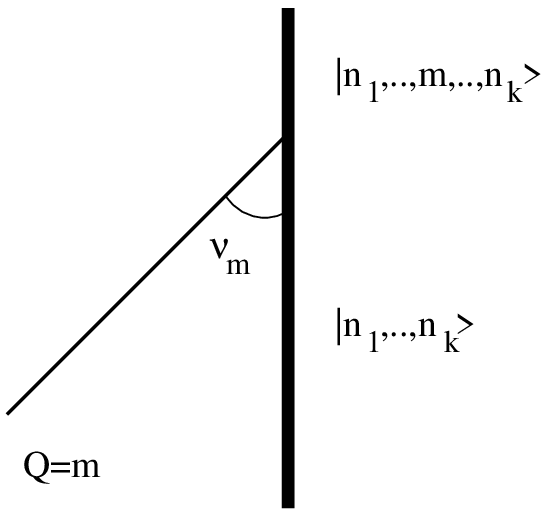}}} 
\caption{: Initial processes}\label{fig:init}\end{figure}\par}
%\vspace{0.3cm}

\begin{figure}\subfigure[Breather
decay]{\resizebox*{!}{5cm}{\includegraphics{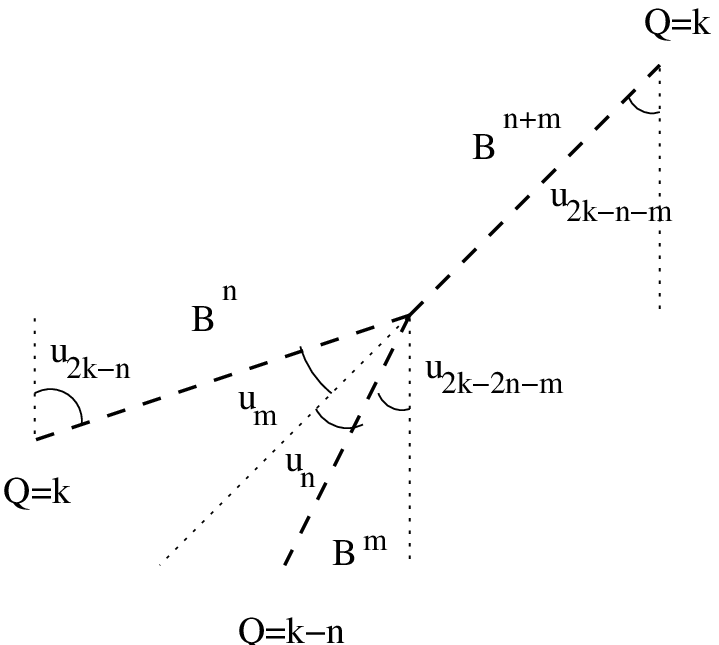}}}
~~~~~~~\subfigure[Breather
fusion]{\resizebox*{!}{5cm}{\includegraphics{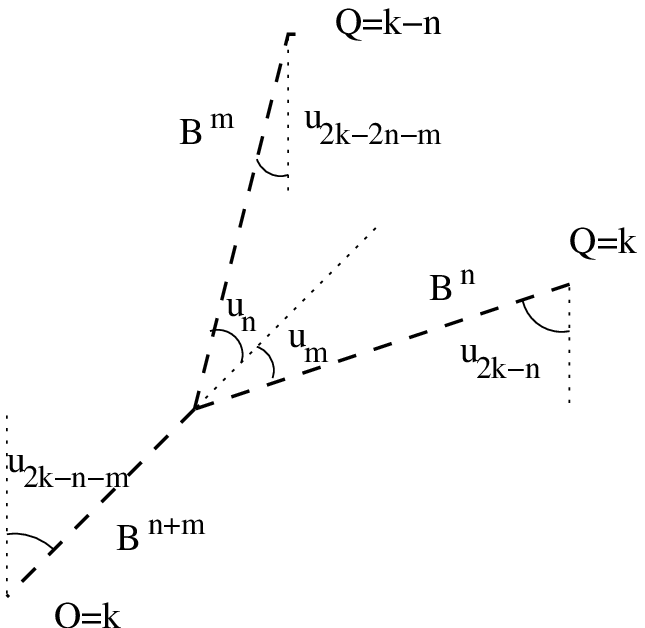}}}
~~~~~~~\subfigure[Wrong
breather]{\resizebox*{!}{5cm}{\includegraphics{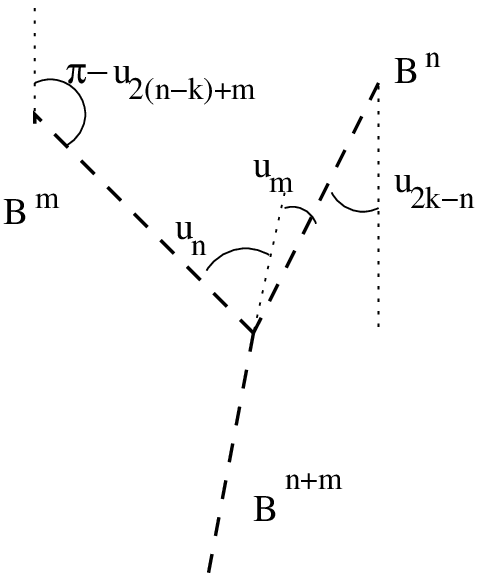}}} 
\caption{}\label{fig:lelegz}\end{figure}

\begin{figure}

%\vspace{0.3cm}
{\centering \subfigure[Breather-soliton]{\resizebox*{!}{5cm}{\includegraphics{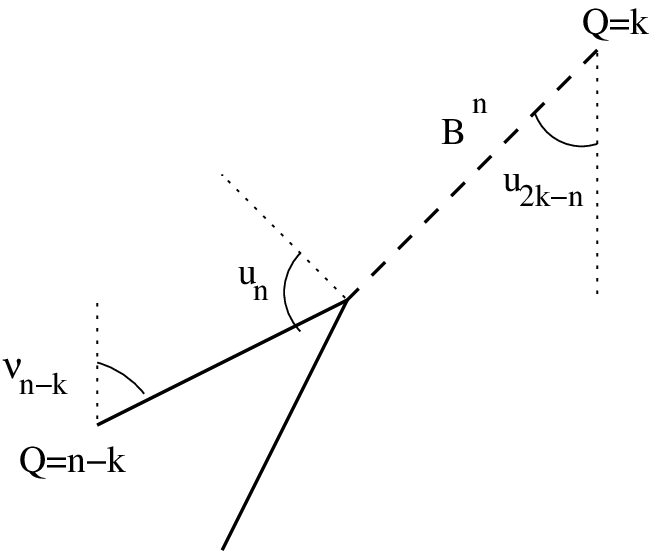}}} ~~\subfigure[Soliton-breather]{\resizebox*{!}{5cm}{\includegraphics{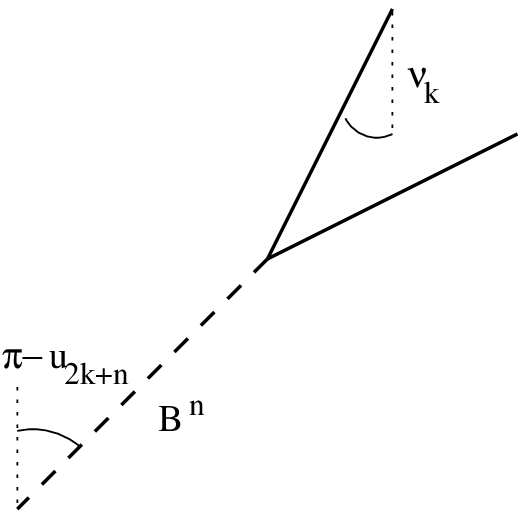}}} ~~~\subfigure[Soliton-soliton]{\resizebox*{!}{4.5cm}{\includegraphics{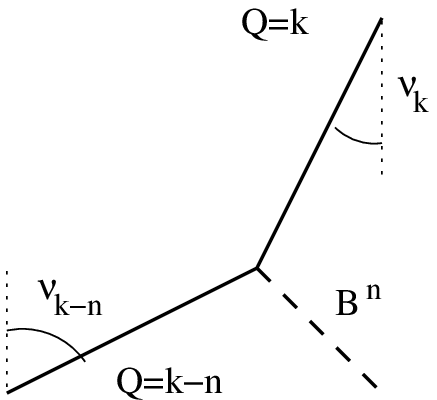}}} \par}
%\vspace{0.3cm}
\caption{}\label{fig:soli}
\end{figure}

We are looking for bulk processes which can take place on the top line
i.e. which are in between the final and initial {}``proper{}'' particles.
There are two types of bulk processes, one with only breathers and
the other one with two solitons and one breather, already depicted
on Fig.(\ref{fig:bulk}). In the present case however, any of the particles can
arrive from the boundary so we have six different cases. First we
analyze how the rapidities are changing. Consider the diagrams on 
Fig.(\ref{fig:lelegz}-\ref{fig:soli}). 

We start with the case when the final particle is a breather. Let
us denote, for a moment, by \( N_{x} \) 
the number of \( x \) type process on the top line,
where \( x\) runs through the six diagrams on Fig.(\ref{fig:lelegz})
and Fig.(\ref{fig:soli}). 
From the conservation of particle types it follows
that in case of a breather initial state \( N_{\rm breather-soliton}=N_{\rm
soliton-breather} \), while in case of a solitonic
initial state \( N_{\rm breather-soliton}+1=N_{\rm soliton-breather}
\) and  all other $N_x$-s are unconstrained.
We claim, however, that only diagrams (a-b) on Fig.(\ref{fig:lelegz}) are allowed.
In showing this we parameterize the rapidities as \( K\frac{\pi }{2}+u_{n} \)
and note that for a {}``proper{}'' breather \( K=0 \), while for
a {}``proper{}'' soliton \( K=-1 \). Now it is easy to see that
on diagrams (a-b) on Fig.(\ref{fig:lelegz}) and on diagram (c) on
Fig.(\ref{fig:soli}) the parameter \( K \)  
does not change, on diagram
(c) on Fig.(\ref{fig:lelegz}) \( K\to K+2 \), on diagram (a) on Fig.(\ref{fig:soli})  
\( K\to K-1 \), finally on 
diagram (b) on Fig.(\ref{fig:soli})  \( K\to K+3 \). Since the initial and final particles
are {}``proper{}'' the various \( K \)-factors must sum up accordingly
on the top line, showing that only diagrams (a-b) on Fig.(\ref{fig:lelegz})  are allowed. We call
these processes {}``proper{}''. 

We also indicated on the diagrams how the charge changes. If we did
not have diagram (b) on Fig.(\ref{fig:lelegz})  we would be ready, since in the
process on diagram (a) the charge never increases, consequently the
final charge can not be larger than the initial one. In the general
case, having also the breather fusion type process we argue as follows:
find the breather fusion vertex on the top line which is farthest away from the
wall. The bottom breather coming from the direction of the wall in this process is also
 a {}``proper{}'' breather. Following its line back towards the wall always turning
to the left we either reach the top line or the wall. Thus this new
line has one or two common vertices with the top line. Now we change
from the original top line to a new {}``top{}'' line by replacing 
the segment of the the old
one between the common
vertices or between the common vertex and the wall
with the line obtained from the bottom breather. Since the new
line segment shares the properties of the original top line, namely
it starts and ends with a {}``proper{}'' breather and we have turned
at each vertex to the left, we can argue for it in the same way as
before. Applying this procedure step by step we arrive at a line which
contains only breather decay type processes so the previous argument applies.
Since in every step we eliminate at least one edge of the diagram
the procedure terminates and we arrive at the final breather concluding
that its charge can not be larger than the largest initial charge
i.e. \( n_{l} \). 

For the solitonic final state the \( K \) parameter is \( -1 \).
Thus either we have purely solitonic processes on the top line, in
which case the charge decreases with each step or we start with some
breather processes allowed in the above mentioned sense and later
by the breather-soliton diagram on Fig.(\ref{fig:soli}) we change back
from the breather to soliton and
continue with the purely solitonic processes. Now in the second case
the previous proof for the breathers can be used to show that the charge
does not increase in the breather part of the top line. In the change
from a breather to soliton the charge always decreases, since for
$B^n$ with charge $k$ the inequality $n<2k$ is always satisfied, 
so the conclusion
is the same i.e. the final charge is always less than the maximal
initial charge \( n_{l} \).  

Summarizing we showed that no on-shell diagram exists, since for the
existence the final charge must be at least \( k \), but the initial
charge is at most \( n_{l}<k \) and the charge does not increase
in the allowed processes. 

These considerations apply to the direct channel process only. The proof for the
crossed channel is quite similar so we omit it here. The difference
is just that we have to introduce a new charge for \( B^{n} \) which
is \( q=\frac{n-m}{2} \) if the rapidity is \( u=u_{m} \). Having
analyzed the analogous diagrams step by step the conclusion is the
same.
 
\section{Pole structure}

In this Appendix we start by reviewing the soliton and breather reflection
factors on the state \( |n_{1},n_{2},\dots ,n_{k}\rangle  \) and
analyze their pole structure. Proceeding inductively in the order
of the pole we show nested on shell diagrams which have poles of the
same order as the reflection amplitudes. We do not compute the residue
of these diagrams, however, since on the one hand the calculation
is very difficult to perform, and on the other, we know instances
where the result does not coincide with the one coming from the reflection
factor. The analysis is carried out in two steps. First, the solitonic
poles are explained, then we turn to the explanation of the breather
ones.

\subsection{Solitonic pole structure}

The solitonic reflection factors are \[
P_{|n_{1},n_{2},\dots ,n_{k}\rangle }(u)=a_{n_{1}}(u)a_{n_{2}}(u)\dots a_{n_{k}}(u)P(u)\quad ;\quad P\leftrightarrow Q\, \, \, .\]
The function \( a_{n}(u)=\{2n-1+\lambda \}\{2n-3+\lambda \}\dots \{1+\lambda \} \)
has the pole structure shown on Fig.(\ref{fig:nupol}), where the number of dots
represents the order of the pole. 

\vspace{0.3cm}
{\centering
\begin{figure}{~~~~~~~~~~~~~~~~~~~~~~~~~~~\resizebox*{8cm}{!}{\includegraphics{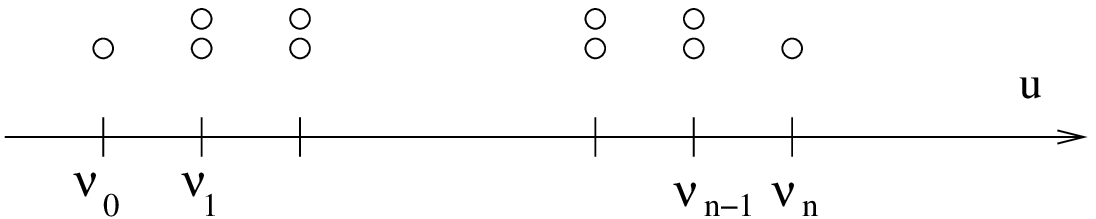}}}
\caption{: Poles of $a_n(u)$}\label{fig:nupol} \end{figure}\par}
\vspace{0.3cm}

In the general reflection amplitude we have to sum up all the orders
of the poles of the various \( a_{n_{i}}(u) \) factors and also to
take into account the poles of the prefactor \( P(u) \). As a result,
we have the following table 

\vspace{0.3cm}
{\centering \begin{tabular}{|c|c|c|c|c|c|c|c|c|c|c|c|}
\hline 
location&
\( \nu _{1} \)&
\( \dots  \)&
\( \nu _{n_{1}-1} \)&
\( \nu _{n_{1}} \)&
\( \nu _{n_{1}+1} \)&
\( \dots  \)&
\( \nu _{n_{k}-1} \)&
\( \nu _{n_{k}} \)&
\( \nu _{n_{k}+1} \)&
\( \dots  \)&
\( \nu _{n_{max}} \)\\
\hline
\hline 
order&
\( 2k+1 \)&
\( \dots  \)&
\( 2k+1 \)&
\( 2k \)&
\( 2k-1 \)&
\( \dots  \)&
\( 3 \)&
\( 2 \)&
\( 1 \)&
\( \dots  \)&
\( 1 \)\\
\hline
\end{tabular}\par}
\vspace{0.3cm}

Now the main observation, similar to what was made in \cite{patr},
is that the poles above are already present in the reflection amplitude
\( P_{|n_{k}\rangle }(u) \), only the orders are different, but this
can be compensated by nesting appropriate sub-diagrams as we will see.
Let us consider the poles at \( \nu _{N} \) step by step.  

\begin{itemize}
\item As we proved in the previous Appendix the simple pole for \( N>n_{k} \)
is responsible for the creation of the state \( |n_{1},n_{2},\dots ,n_{k},N\rangle  \). 
\item Lets consider diagram (a) on Fig.(\ref{fig:1sol2br}). The state \( |n_{1},n_{2},\dots ,n_{k-1},n_{k}\rangle  \)
decays to \newline \( |n_{1},n_{2},\dots ,n_{k-1}\rangle  \) by emitting an
anti soliton with rapidity \( \nu _{n_{k}} \). The soliton and the
anti soliton fuse into \( B^{N+n_{k}} \) with rapidity \( u_{n_{k}-N} \).
In the \( N=n_{k} \) case the breather travels parallel with the
wall giving a diagram with a second order pole.  
\item In the \( N<n_{k} \) case the reflection factor of \( B^{N+n_{k}} \)
has a pole at \( u=u_{n_{k}-N} \), for which a sub-diagram, that is
described in detail in the next subsection, has to be embedded.  We
just present here in advance, that when \( N \) decreasingly reaches
any of the \( n_{i} \)-s then the order of the breather pole is increased
by one and when \( N=n_{i}-1 \) it increases by one more. Furthermore,
it does not change until \( N \) reaches \( n_{i-1} \), where the
same story happens. Collecting all the order of these poles and also
the one more extra, which comes from the ground state reflection factor
we arrive at the right result. 
\end{itemize}
We emphasize that from the shifting argument described in section  
(3.3) we know that in spite of the existence of the on-shell diagram
shown above the pole at \( \nu _{N} \) in the reflection amplitude
\( P_{|n_{1},n_{2},\dots ,n_{k}\rangle }(u) \) also creates the state
\( |n_{1},\dots ,n_{i},N,n_{i+1},\dots ,n_{k}\rangle  \), if this
state exists, that is if \( n_{i}<N<n_{i+1} \) for some \( i \).
 
\subsection{Breather pole structure}

The breather \( B^{n} \) on the state \( |n_{1},n_{2},\dots ,n_{k}\rangle  \)
has the following reflection factor \[
R^{(n)}_{|n_{1},n_{2},\dots ,n_{k}\rangle }(u)=b^{n}_{n_{1}}(u)b^{n}_{n_{2}}(u)\dots b^{n}_{n_{k}}(u)R^{(n)}(u)\]
where \[
b^{n}_{m}(u)=\{n+2m-1\}\{n+2m-3\}\dots \{n+2m-1-2l\}\dots \left\{ \begin{array}{c}
\{n-2m+1\}\quad \textrm{if}\quad n-2m+1>0\\
\{2m-n+1\}\quad \textrm{if}\quad 2m-n+1>0
\end{array}\ .\right. \]
Lets parameterize the poles of \( b_{m}^{n}(u) \) as \( u_{n+2x} \).
Considering the two cases separately we obtain the location of the
poles as 

\vspace{0.3cm}
{\centering \resizebox*{8cm}{!}{\includegraphics{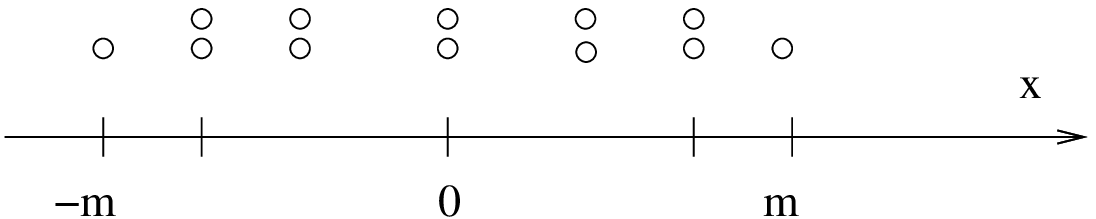}}  \par}
\vspace{0.3cm}
in the case when \( n>2m \) while as 

\vspace{0.3cm}
{\centering \resizebox*{8cm}{!}{\includegraphics{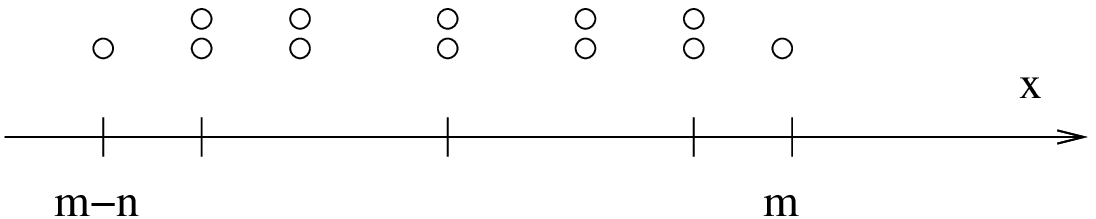}}  \par}
\vspace{0.3cm}
in the other case. Summing up all the poles coming from the various
\( b \) -factors we get the following result. If \( n_{q}\leq x<n_{q+1} \)
and \( n_{p}\leq n+x<n_{p+1} \) hold in the \( x>0 \) case or \( n_{q}\leq |x|<n_{q+1} \)
and \( n_{p}<n-|x|<n_{p+1} \) is satisfied for \( x<0 \), then the
order of the pole at \( u_{n+2x} \) is \( 2(p-q)+\epsilon  \) (where
\( \epsilon =1 \) if the bound for \( p \) is saturated and \( \epsilon =-1 \)
if the bound for \( q \) is saturated and zero otherwise). For each
pole there is the analogous pole at \( \pi -u_{n+2x} \), which is
in the physical strip if \( u_{n+2x} \) is not there. 

Now we explain the pole of the reflection amplitude \( R^{(n)}_{|n_{1},n_{2},\dots ,n_{k}\rangle }(u) \)
at \( u_{n+2x} \). For \( x=n_{k} \) the state \( |n_{1},n_{2},\dots ,n_{k}+n\rangle  \)
is created or if it is not in the spectrum we have an analogous diagram
to diagram (d) on Fig.(\ref{fig:bound}). For \( x<n_{k} \) consider diagram
(a) on Fig.(\ref{fig:2leleg}). 

\vspace{0.3cm}
{\centering \begin{figure}~~~~~~~~~\subfigure[Breather decay]{\resizebox*{!}{6cm}{\includegraphics{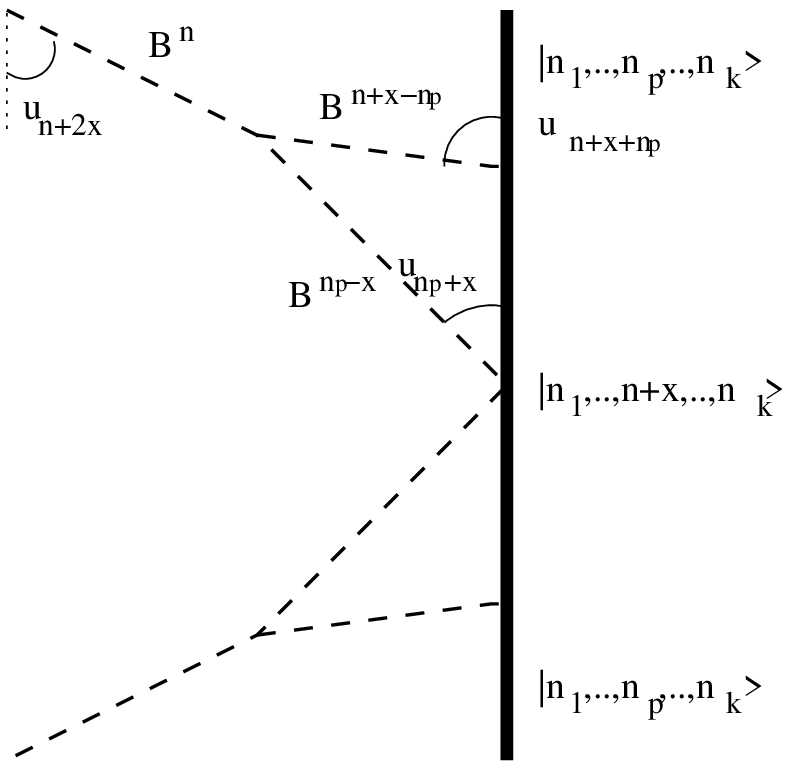}}} ~~~~~\subfigure[Soliton decay]{\resizebox*{!}{6cm}{\includegraphics{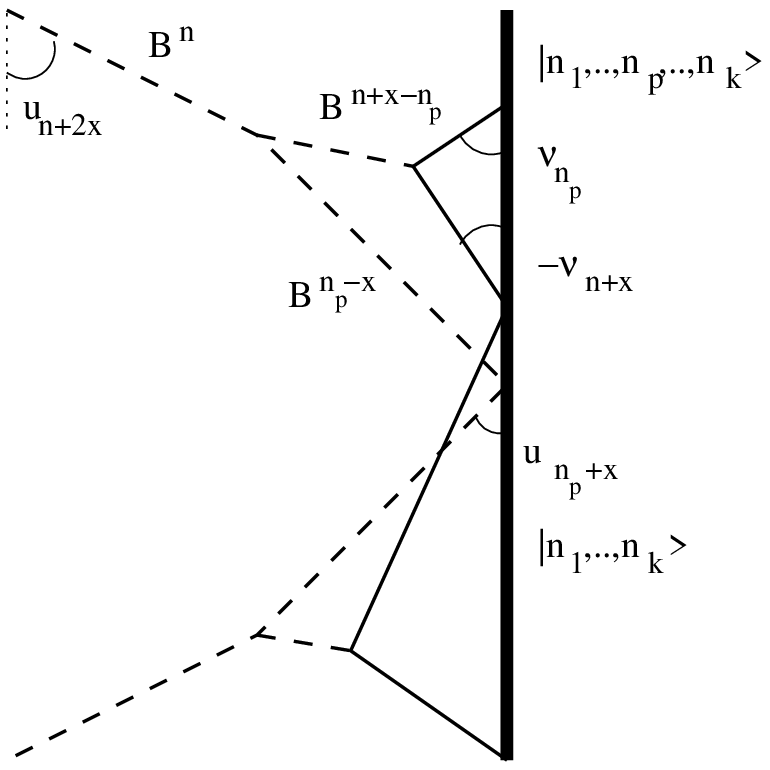}}} \caption{}\label{fig:2leleg}\end{figure}\par}
%\vspace{0.3cm}

Here \( B^{n} \) decays into \( B^{n+x-n_{p}} \) and \( B^{n_{p}-x} \)
in the bulk. First \( B^{n+x-n_{p}} \) reaches the wall with rapidity
\( u_{n+x+n_{p}} \) exciting the state \( |n_{1},\dots ,n_{p},\dots ,n_{k}\rangle  \)
to \( |n_{1},\dots ,n+x,\dots ,n_{k}\rangle  \). If the breather
traveled backwards in time (\( n+x+n_{p}>[\lambda ] \)) or the excited
state did not exists (\( n+x>\left[ \frac{\lambda }{2}\right]  \))
then we would have diagram (b), where the state \( |n_{1},\dots ,n_{p-1},n_{p},n_{p+1},\dots ,n_{k}\rangle  \)
would decay to \( |n_{1},\dots ,n_{p-1},n_{p+1},\dots ,n_{k}\rangle  \)
by the emission of a soliton with rapidity \( -\nu _{n_{p}} \), which
then absorbs \( B^{n+x-n_{p}} \) and reaches the wall at rapidity
\( -\nu _{n+x} \) . In both cases \( B^{n_{p}-x} \) reflects on
the resulting state with rapidity \( u_{(n_{p}-x)+2x} \). In order
to compute the order of its pole we observe that \( n_{q}\leq x<n_{q+1} \)
as before, but now \( n_{p-1}<(n_{p}-x)+x \). From this it follows
that the order of the pole is \( 2(p-1-q)+\epsilon  \), which is
less than the one investigated originally by two. Since the diagram
itself gives a second order pole (we have used the Coleman-Thun type
cancellation in the (b) case) we are ready. Resolving inductively
the pole of this reflection by the analogous sub-diagram we arrive
at cascades of diagrams, where in the last sub-diagram the breather
reflects with a simple pole or without poles. 

For the poles of type \( \pi -u_{n+2x} \) we have an analogous consideration:
If we parameterize the rapidity as \( \pi -u_{n+2x} \) and analyze
the relation between \( x,n+x \) and \( n_{i} \) we get the same
formula for the order of the poles as we had before. The explanation
is given in terms of the crossed diagrams. Diagram (b) on
Fig.(\ref{fig:1sol2br}) 
explains the \( x=n_{k} \) case, while for \( x<n_{k} \) we have
diagram (c) on the same Figure.
The argument is completely analogous to the non-crossed version so
we omit it here. 

\vspace{0.3cm}
{\centering \begin{figure}~\subfigure[Soliton
decay]{\resizebox*{!}{6cm}{\includegraphics{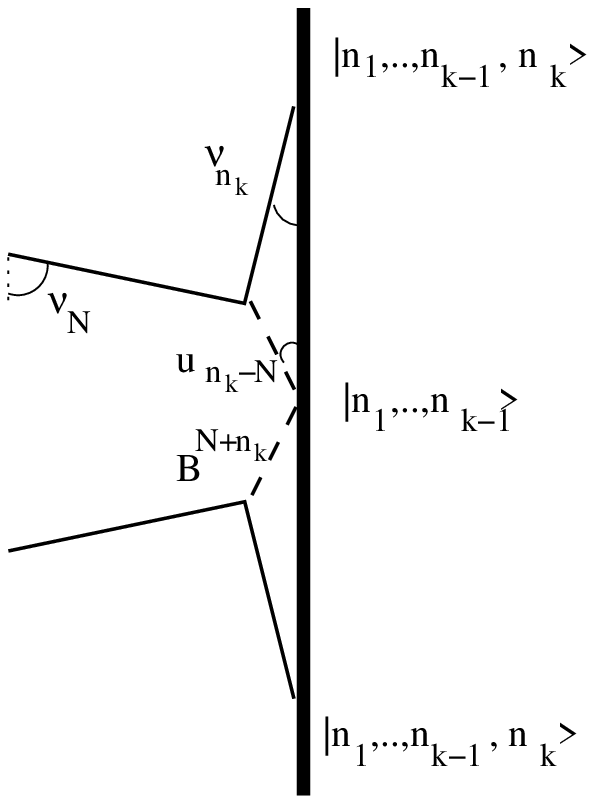}}}
~~\subfigure[Soliton decay crossed]{\resizebox*{!}{6cm}{\includegraphics{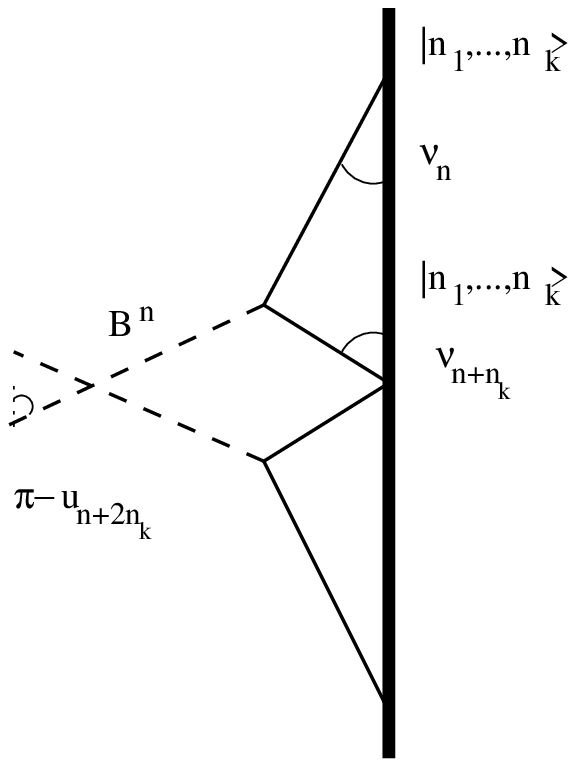}}} ~~~~
\subfigure[Breather
decay crossed]{\resizebox*{!}{6cm}{\includegraphics{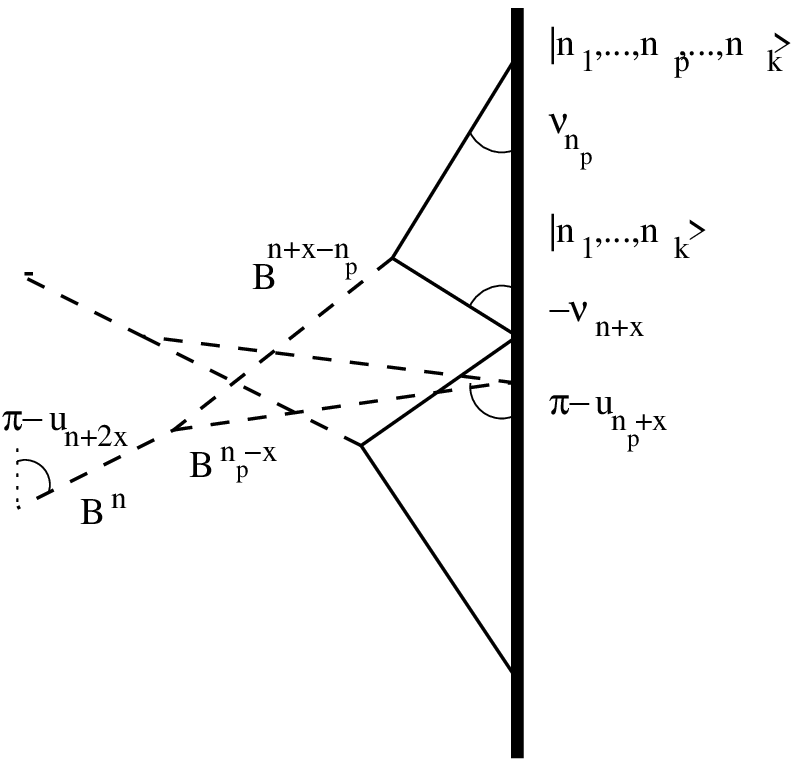}}} 
\caption{}\label{fig:1sol2br}\end{figure}\par}
\vspace{0.3cm}
Note, that these breather diagrams are generic, which means that they
exist for any \( x \). It may happen, however, that for some particular
\( x \) we have some other diagram with the right order of pole.
Notice also that in spite of the existence of the diagrams above we
might have boundary bound state creation, described in detail in
section (3.3), at the same time. 

\section{Boundary $c=1$ theories}

In this Appendix we warm up by considering the compactified free boson
with periodic boundary conditions 
on a circle of circumference \( L \). Having mapped the system onto the plane
we identify those conformally invariant boundary conditions which also preserve
the underlying affine \( \hat{U}_{1}\times \hat{U}_{1} \) symmetry. They originate
from cutting the original circle and applying Neumann or Dirichlet boundary
conditions on both ends of the strip. We also determine the theory
which corresponds to imposing Neumann boundary condition on one end and Dirichlet
boundary condition on the other end of the strip. In all cases the data we need
for the TCSA, such as the spectrum of bulk and boundary primary operators, their
normalizations, conformal weights, bulk and boundary operator product expansions
are summarized. Though most of these data are available in the
literature we collect them here to make the paper self contained.

\subsection{Compactified free boson with periodic boundary conditions}

Consider a free boson with compactification radius \( r \) confined into a
box of size \( L \) subject to periodic boundary condition \( \Phi (L,t)\equiv \Phi (0,t)+2\pi rn \),
where \( n\in {\mathbb Z}\) is called the winding number. The action which governs the
dynamics is \[
S=\frac{1}{8\pi }\int ^{\infty }_{-\infty }dt\int _{0}^{L}dx\partial _{\mu }\Phi \partial ^{\mu }\Phi \]
 Canonical quantization results in the following expression for the field operator
\( \Phi (x,t) \):\[
\Phi (x,t)=\Phi _{0}+\frac{4\pi }{L}\left( \Pi _{0}t+\frac{rM}{2}x\right) +i\sum _{n\neq 0}\frac{1}{n}\left( a_{n}e^{i\frac{2\pi }{L}n(x-t)}+\bar{a}_{n}e^{-i\frac{2\pi }{L}n(x+t)}\right) ,\]
 where \( M \)
is the winding operator with eigenvalues \( m\in {\mathrm Z}\), and the nonzero commutators are \begin{equation}
\label{u1alg}
[\Phi _{0},\Pi _{0}]=i\quad ;\qquad [a_{n},a_{m}]=n\delta _{n+m}\quad
,\quad [\bar{a}_{n},\bar{a}_{m}]=n\delta _{n+m}\ .
\end{equation}
We can use \( \zeta  \)-function regularization for computing the Casimir energy and
find the Hamiltonian as \[
H=\frac{2\pi }{L}\sum
_{n>0}(a_{-n}a_{n}+\bar{a}_{-n}\bar{a}_{n}-\frac{1}{12})+\frac{2\pi
}{L}\left( \Pi _{0}^{2}+\left( \frac{rM}{2}\right) ^{2}\right)\ . \]
 Having mapped the system onto the plane via the \( (x,it)=\xi \to z=e^{i\frac{2\pi }{L}\xi } \)
conformal transformation, chiral factorization originates 
from the split \( \Phi (z,\bar{z})=\phi (z)+\bar{\phi }(\bar{z}) \),
where \[
\phi (z)=\phi _{0}-ia_{0}\ln z+i\sum _{n\in
{\mathbb Z}}a_{n}\frac{z^{-n}}{n}\quad ;\qquad \bar{\phi }(\bar{z})=\bar{\phi
}_{0}-i\bar{a}_{0}\ln \bar{z}+i\sum _{n\in
{\mathbb Z}}\bar{a}_{n}\frac{\bar{z}^{-n}}{n}\ .\]
It is useful to introduce the dual of \( \Phi (z,\bar{z}) \) as \( \tilde{\Phi }(z,\bar{z})=\phi (z)-\bar{\phi }(\bar{z}) \).
The \( \hat{U}_{1}\times \hat{U}_{1} \) symmetry of the model is generated
by the chiral currents, \( J(z)=i\partial _{z}\phi (z)\, ,\, \, \bar{J}(\bar{z})=i\partial _{\bar{z}}\bar{\phi }(\bar{z}) \),
and the primary fields of the symmetry algebra are the vertex operators\begin{equation}
\label{vop}
V_{(n,m)}(z,\bar{z})=\, :e^{i\frac{n}{r}\Phi (z,\bar{z})+i\frac{mr}{2}\tilde{\Phi }(z,\bar{z})}:=\, :e^{iq\phi (z)+i\bar{q}\bar{\phi }(\bar{z})}:
\end{equation}
We use the parameterization \( (n,m) \) and \( (q,\bar{q}) \) in
parallel, 
the connection between them being \( q+\bar{q}=\frac{2n}{r} \) and \( q-\bar{q}=mr \).
The vertex operators have conformal weights \( h_{(n,m)}=\frac{q^{2}}{2}\, ,\, \, \bar{h}_{(n,m)}=\frac{\bar{q}^{2}}{2} \)
with respect to the conformal energy momentum tensor \( T(z)=\frac{1}{2}:J(z)J(z): \)
and \( \bar{T}(z)=\frac{1}{2}:\bar{J}(z)\bar{J}(z): \). The Hilbert space is
built up by the successive application of the modes of the chiral currents on
the highest weight state created by the vertex operators as\[
|n,m\rangle =V_{(n,m)}(0,0)|0\rangle \]
 and \( \Pi _{0}|n,m\rangle =n|n,m\rangle \, ,\, \, M|n,m\rangle =m|n,m\rangle  \).
The operator product expansion of the vertex operators starts as \begin{equation}
\label{vope}
V_{(q,\bar{q})}(z,\bar{z})V_{(q^{'},\bar{q}^{'})}(w,\bar{w})=(z-w)^{qq^{'}}(\bar{z}-\bar{w})^{\bar{q}\bar{q}^{'}}V_{(q+q^{'},\bar{q}+\bar{q}^{'})}(w,\bar{w})+\dots 
\end{equation}
 We have two maximal sets of the allowed \( (n,m) \) pairs \cite{KM} and here
we concentrate on the bosonic one which has \( n\in {\mathbb Z},m\in {\mathbb Z} \).

\subsection{Conformal boundary conditions}

Now we would like to impose such boundary conditions in the theory that preserve
not only the conformal but also the Kac-Moody symmetry of the model. The easiest
way to formulate this is to restrict the theory on the upper half plane and
demand\cite{Zuber} \[
L(z)=\bar{L}(\bar{z})\quad ,\quad J(z)=\Omega \bar{J}(\bar{z})\quad ;\]
 on the real axis, where \( \Omega  \) is any automorphism of the algebra (\ref{u1alg}).
We have two possible choices \( \Omega =1 \) or \( \Omega =-1 \), \cite{RSch},
they correspond to Neumann and Dirichlet boundary conditions imposed on the
field \( \Phi  \), respectively. Note that for the dual field the roles are
interchanged, that is \( \Omega =1 \) corresponds to Dirichlet while \( \Omega =-1 \)
for Neumann boundary condition.

\subsubsection{Neumann boundary condition}

The \( \Omega =1 \) boundary condition in terms of the modes reads as \( \bar{a}_{n}=a_{n} \),
that is \[
\Phi (z,\bar{z})=\Phi _{0}-ia_{0}(\ln z+\ln \bar{z})+i\sum _{n\in
{\mathbb Z}}\frac{a_{n}}{n}(z^{-n}+\bar{z}^{-n})\ .\]
 We can map this field back to the strip by the inverse of the transformation
\( \xi \to z=e^{i\frac{\pi }{L}\xi } \) \begin{equation} \label{Nfield}
\Phi (x,t)=\Phi _{0}+\frac{4\pi }{L}\Pi _{0}+\sum _{n\neq 0}\frac{a_{n}}{n}(e^{i\frac{\pi }{L}n(x-t)}+e^{-i\frac{\pi }{L}n(x+t)})\end{equation}
 From the \( x \) -dependence we can read off the Neumann boundary conditions
\[
\partial _{x}\Phi (0,t)=\partial _{x}\Phi (L,t)=0\quad ;\quad \forall t\]
 The Hamiltonian of this system in the strip can be obtained directly from (\ref{Nfield})
by \( \zeta  \)-function regularization\begin{equation} \label{X}
H=\frac{2\pi }{L}\Pi _{0}^{2}+\frac{\pi }{L}\left( \sum _{n\neq
 0}na_{-n}a_{n}-\frac{1}{24}\right)\ . \end{equation}
 Turning back to the plane we realize that the vertex operators \( V_{(n,0)}(z,\bar{z}) \)
do not reproduce the most singular part of the OPE (\ref{vope}) any
 more as we would expect. 
Changing however
their normalization function as\begin{equation} \label{X1}
V_{(n,0)}(z,\bar{z})=|z-\bar{z}|^{q^{2}}\, :e^{iq\Phi \phi
 (z,\bar{z})}:\quad ;\quad q=\frac{n}{r}\ ,\end{equation}
 we preserve their conformal weights, (which is \( h=\frac{q^{2}}{2} \) for \( T(z) \))
and restore the leading part of the OPE (\ref{vope}). The boundary fields in a boundary
conformal field theory live only on the boundary  and are in a one-to-one correspondence
with the vectors of the Hilbert space. In our case they are the boundary vertex
operators\[
\Psi _{q}(x)=\, :e^{iq\Phi (x,x)}:\quad ;\quad q=\frac{n}{r}\ ,\]
(which are primary fields of weight \( h_{n}=2q^{2} \)), and their descendants.
The dual field vanishes at the boundary so in the Hilbert space we do not have
winding as we would expect from the Neumann boundary condition.

\subsubsection{Dirichlet boundary condition}

The \( \Omega =-1 \) boundary condition forces \( \Phi (z,\bar{z}) \) to be
of the form\[
\Phi (z,\bar{z})=\Phi _{0}-ia_{0}(\ln z-\ln \bar{z})+i\sum _{n\in
{\mathbb Z}}\frac{a_{n}}{n}(z^{-n}-\bar{z}^{-n})\ .\]
 The boundary condition can be determined on the strip by mapping the system
via the inverse of \( \xi \to z=e^{i\frac{\pi }{L}\xi } \) to the strip. We
have\[
\Phi (x,t)=\Phi _{0}+\frac{2\pi }{L}r\tilde{M}x-i\sum _{n\neq
0}\frac{a_{n}}{n}\left( e^{i\frac{\pi }{L}n(x-t)}-e^{-i\frac{\pi
}{L}n(x+t)}\right)\ , \]
 where \( \tilde{M}=M+\frac{\Phi _{L}-\Phi _{0}}{2\pi r} \) and \( M\in {\mathbb Z} \)
is the winding number. This corresponds to the \[
\Phi (0,t)=\Phi _{0}\quad ;\quad \Phi (L,t)=\Phi _{L}\quad \forall t\]
Dirichlet boundary condition. The Hamiltonian of the system above is \begin{equation}\label{XX}
H=\frac{2\pi }{L}\left( \frac{r\tilde{M}}{2}\right) ^{2}+\frac{\pi
}{L}\left( \sum _{n\neq 0}na_{-n}a_{n}-\frac{1}{24}\right)\ . \end{equation}
By changing the normalization function of the vertex operators 
\begin{equation}\label{XX1}
V_{(q,0)}(z,\bar{z})=|z-\bar{z}|^{-q^{2}}\, :e^{iq\Phi \phi
(z,\bar{z})}:\quad ;\quad q=\frac{n}{r}\ ,\end{equation}
we can ensure the most singular part of 
the bulk OPE to hold. Now the field \( \Phi  \) is constant at
the boundary so we do not have nonzero momentum Fock modules. The dual field
however can create the sectors corresponding to the different winding numbers.
The Hilbert space by this token consists of the Fock modules corresponding to
the different winding numbers.

\subsubsection{Mixed boundary condition}

It is also possible to demand mixed boundary conditions. On the plane one possible
choice is Dirichlet boundary condition for \( \Im m (z)=0\, ,\, \, \Re e (z)>0 \)
and Neumann for \( \Im m (z)=0\, ,\, \, \Re e (z)<0. \) A field satisfying
this can be given 
as \[
\Phi (z,\bar{z})=\Phi _{0}+i\sum _{n\in
{\mathbb Z}+\frac{1}{2}}\frac{a_{n}}{n}(z^{-n}-\bar{z}^{-n})\ .\]
 Since on the strip \[
\Phi (x,t)=\Phi _{0}-i\sum _{n\in {\mathbb Z}+\frac{1}{2}}\frac{a_{n}}{n}\left(
e^{i\frac{\pi }{L}n(x-t)}-e^{-i\frac{\pi }{L}n(x+t)}\right)\ , \]
the boundary condition there reads as
 \[
\Phi (0,t)=\Phi _{0}\quad ;\quad \partial _{x}\Phi (L,t)=0\quad
\forall t\ ,\]
that is Dirichlet boundary condition at \( x=0 \) and Neumann boundary condition
at \( x=L \). Using \( \zeta  \)-function regularization the Hamiltonian turns out to
be \begin{equation}\label{3X}
H=\frac{\pi }{L}\left( \sum _{n\in
{\mathbb Z}+\frac{1}{2}}na_{-n}a_{n}+\frac{1}{48}\right)\ . \end{equation}
The model looks like a twisted boson with chiral current \(
J(z)=\sum\limits_{n\in {\mathbb Z}+\frac{1}{2}}a_{n}z^{-n-1}\ .\)
The vertex operators acquire a factor \begin{equation}\label{3X1}
V_{q}(z,\bar{z})=\left(
\frac{\sqrt{\frac{z}{\bar{z}}}+2+\sqrt{\frac{\bar{z}}{z}}}{4|z-\bar{z}|}\right)
^{q^{2}}\, :e^{iq\Phi  (z,\bar{z})}:\quad ;\quad q=\frac{n}{r}\ ,\end{equation}
in order to ensure the most singular part of 
the bulk OPE to hold. Now the field \( \Phi  \) has a non vanishing
limit only for \( \Im m (z)=0\, ,\, \, \Re e (z)<0 \) while the dual field survives
at \( \Im m (z)=0\, ,\, \, \Re e (z)>0 \). Thus the only boundary operator is the
current \( J(x) \) and consequently the Hilbert space is generated by acting with
its modes \( a_{-n}\, ,\, \, n\in {\mathbb Z}_{+}-\frac{1}{2} \), on the vacuum vector.

\end{document}